\documentclass[sigconf]{acmart}

\usepackage{booktabs}
\usepackage{graphicx}
\usepackage{subfig}
\usepackage{multirow}
\usepackage{balance}
\usepackage{url,graphicx}
\usepackage{tabularx}
\usepackage{mathrsfs, amsfonts}
\usepackage{multirow}
\usepackage{booktabs}
\usepackage{longtable}
\usepackage{colortbl}
\usepackage{xcolor}
\usepackage{mdwlist}
\usepackage{marvosym}
\usepackage{bm}
\usepackage{amsmath}
\usepackage{dsfont}
\usepackage{comment}
\usepackage{amssymb}
\usepackage{latexsym}
\usepackage{listings}
\usepackage[ruled,linesnumbered]{algorithm2e}
\usepackage{paralist}
\usepackage{enumitem}
\usepackage{threeparttable}
\setlist{nosep}

\usepackage{soul}
\graphicspath{{eps/}}
\usepackage{setspace}

\usepackage[np, autolanguage]{numprint}

\newcommand{\hidethis}[1]{}

\allowdisplaybreaks

\makeatletter
\renewcommand{\paragraph}[1]{\medskip\noindent\emph{#1.~}}
\makeatother

\copyrightyear{2017}
\acmYear{2017}
\setcopyright{acmcopyright}
\acmConference{CIKM'17 }{}{November 6--10, 2017, Singapore.}
\acmPrice{15.00}
\acmDOI{https://doi.org/10.1145/3132847.3132926}
\acmISBN{ISBN 978-1-4503-4918-5/17/11}

\begin{document}

\title{Neural Attentive Session-based Recommendation}

\author{Jing Li}
\affiliation{
  \institution{Shandong University}
  \city{Jinan} 
  \state{China} 
}
\email{jingli.sdu@gmail.com}

\author{Pengjie Ren}
\affiliation{
  \institution{Shandong University}
  \city{Jinan} 
  \state{China} 
}
\email{jay.ren@outlook.com}

\author{Zhumin Chen}
\affiliation{
	\institution{Shandong University}
	\city{Jinan} 
	\state{China} 
}
\email{chenzhumin@sdu.edu.cn}

\author{Zhaochun Ren}
\affiliation{
	\institution{Data Science Lab, JD.com}
	\city{Beijing} 
	\state{China} 
}
\email{renzhaochun@jd.com}

\author{Tao Lian}
\affiliation{
	\institution{Shandong University}
	\city{Jinan} 
	\state{China} 
}
\email{liantao1988@gmail.com}

\author{Jun Ma}
\affiliation{
	\institution{Shandong University}
	\city{Jinan} 
	\state{China} 
}
\email{majun@sdu.edu.cn}

\renewcommand{\shortauthors}{J. Li et al.}

\begin{abstract}

Given e-commerce scenarios that user profiles are invisible, session-based recommendation is proposed to generate recommendation results from short sessions.
Previous work only considers the user's sequential behavior in the current session, whereas the user's main purpose in the current session is not emphasized.
In this paper, we propose a novel neural networks framework, i.e., Neural Attentive Recommendation Machine (NARM), to tackle this problem. Specifically, we explore a hybrid encoder with an attention mechanism to model the user's sequential behavior and capture the user's main purpose in the current session, which are combined as a unified session representation later. We then compute the recommendation scores for each candidate item with a bi-linear matching scheme based on this unified session representation. We train NARM by jointly learning the item and session representations as well as their matchings. 
We carried out extensive experiments on two benchmark datasets. Our experimental results show that NARM outperforms state-of-the-art baselines on both datasets. Furthermore, we also find that NARM achieves a significant improvement on long sessions, which demonstrates its advantages in modeling the user's sequential behavior and main purpose simultaneously.

\end{abstract}

%\begin{CCSXML}
%	<ccs2012>
%	<concept>
%	<concept_id>10002951.10003317.10003347.10003350</concept_id>
%	<concept_desc>Information systems~Recommender systems</concept_desc>
%	<concept_significance>500</concept_significance>
%	</concept>
%	<concept>
%	<concept_id>10010147.10010257.10010258.10010259</concept_id>
%	<concept_desc>Computing methodologies~Supervised learning</concept_desc>
%	<concept_significance>500</concept_significance>
%	</concept>
%	<concept>
%	<concept_id>10010147.10010257.10010293.10010294</concept_id>
%	<concept_desc>Computing methodologies~Neural networks</concept_desc>
%	<concept_significance>500</concept_significance>
%	</concept>
%	</ccs2012>
%\end{CCSXML}

%\ccsdesc[500]{Information systems~Recommender systems}
%\ccsdesc[500]{Computing methodologies~Supervised learning}
%\ccsdesc[500]{Computing methodologies~Neural networks}

\keywords{Session-based recommendation, sequential behavior, recurrent neural networks, attention mechanism}

\maketitle

\section{INTRODUCTION}

A user session is kicked off when a user clicks a certain item; within a user session,  
clicking on the interesting item, and spending more time viewing it. After that, the user clicks another interesting one to start the view again. Such iterative process will be completed until the user's requirements are satisfied. 
Current recommendation research confronts challenges when recommendations are merely from those user sessions, where existing recommendation methods \cite{koren2009matrix,adomavicius2005toward,weimer2007maximum,su2009survey} cannot perform well. To tackle this problem, session-based recommendation \cite{schafer1999recommender} is proposed to predict the next item that the user is probably interested in based merely on implicit feedbacks, i.e., user clicks, in the current session.

    \begin{figure}
    	\centering
    	\vspace{1em}
    	\subfloat[The global recommender]{
    		\includegraphics[height=0.6in, width=3.1in]{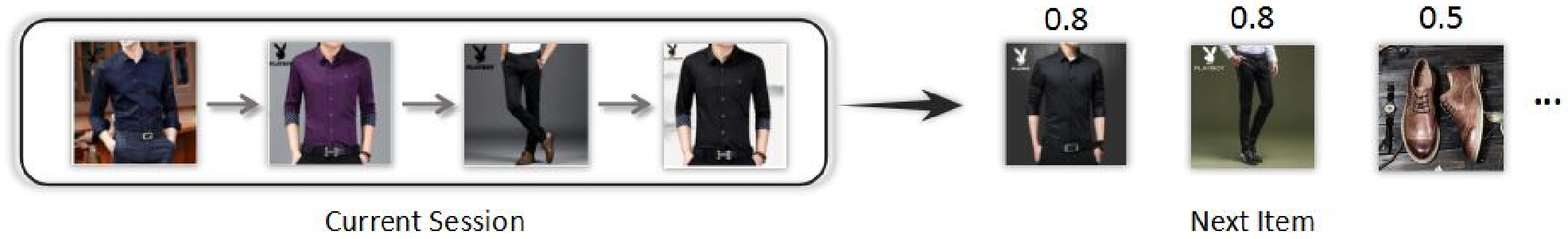}}
    	\hspace{4em} 
    	\subfloat[The local recommender]{
    		\includegraphics[height=0.6in, width=3.1in]{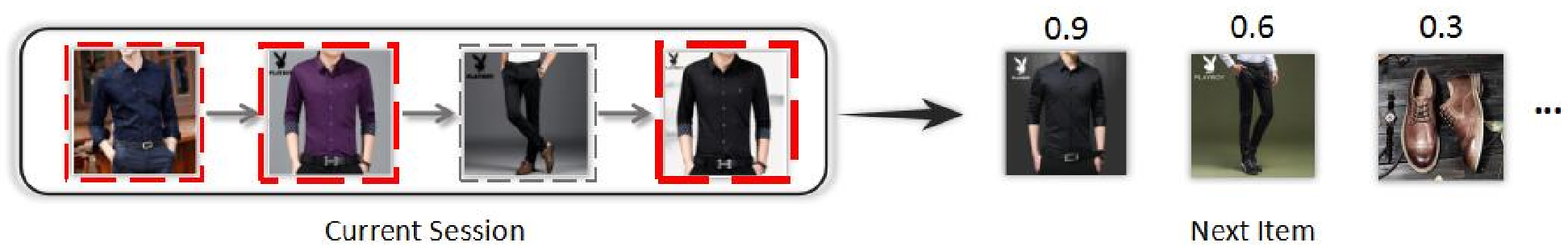}}	
    	\caption{Two different recommenders. The global recommender models the user's whole sequential behavior to make recommendations while the local recommender captures the user's main purpose to make recommendations. The numbers above the items denote the recommendation scores produced by each recommender. In (b), the item in the red dashed box is more relevant to the current user's intention. And the red line is thicker when the item is more important.}
    \end{figure}

\citet{hidasi2015session} apply recurrent neural networks (RNN) with Gated Recurrent Units (GRU) for session-based recommendation. The model considers the first item clicked by a user as the initial input of RNN, and generates recommendations based on it. Then the user might click one of the recommendations, which is fed into RNN next, and the successive recommendations are produced based on the whole previous clicks. \citet{tan2016improved} further improve this RNN-based model by utilizing two crucial techniques, i.e., data augmentation and a method to account for shifts in the input data distribution. Though all above RNN-based methods show promising improvements over traditional recommendation approaches, they only take into account the user's sequential behavior in the current session, whereas the user's main purpose in the current session is not emphasized. Relying only on the user's sequential behavior is dangerous when a user accidentally clicks on wrong items or s/he is attracted by some unrelated items due to curiosity. Therefore, we argue that both the user's sequential behavior and main purpose in the current session should be considered in session-based recommendation.

Suppose that a user wants to buy a shirt on the Internet. As shown in Figure 1, during browsing, s/he tends to click on some shirts with similar styles to make a comparison, meanwhile s/he might click a pair of suit pants by accident or due to curiosity. After that, s/he keeps looking for suitable shirts. In this case, if we only consider about his/her sequential behavior, another shirt or suit pants even a pair of shoes might be recommended because many users click them after clicking some shirts and suit pants, as shown in Figure 1(a). Assume that the recommender is an experienced human purchasing guide, the guide could conjecture that this user is very likely to buy a short sleeve shirt at this time because most of his/her clicked items are related to it. Therefore, more attention would be paid to the short sleeve shirts that the user has clicked and another similar shirt would be recommended, as shown in Figure 1(b). Ideally, in addition to considering about the user's entire sequential behavior, a better recommender should also take into account the user's main purpose which is reflected by some relatively important items in the current session. Note that the sequential behavior and the main purpose in one session are complementary to each other because we can not always conjecture a user's main purpose from a session, e.g., when the session is too short or the user just clicks something aimlessly.

To tackle the above problem, we propose a novel neural networks framework, namely Neural Attentive Recommendation Machine (NARM). Specifically, we explore a hybrid encoder with an attention mechanism to model the user's sequential behavior and capture the user's main purpose in the current session, which are combined as a unified session representation later. With this item-level attention mechanism, NARM learns to attend differentially to more and less important items. We then compute the recommendation scores for each candidate item with a bi-linear matching scheme based on the unified session representation. NARM is trained by jointly learning the item and session representations as well as their matchings.

\noindent%
The main contributions of this work are summarized as follows:
 \begin{itemize}
\item We propose a novel NARM model to take into account both the user's sequential behavior and main purpose in the current session, and compute recommendation scores by using a bi-linear matching scheme. 
%We train NARM by jointly learning the item and session representations as well as their matchings.
\item We apply an attention mechanism to extract the user's main purpose in the current session. 
%This gives NARM the ability to pay more attention to more important clicked items when constructing the current session representation, so that NARM can make a preciser recommendation.
\item We carried out extensive experiments on two benchmark datasets. The results show that NARM outperforms state-of-the-art baselines in terms of recall and MRR on both datasets. Moreover, we find that NARM achieves better performance on long sessions, which demonstrates its advantages in modeling the user's sequential behavior and main purpose simultaneously.
\end{itemize}

\section{RELATED WORK}

Session-based recommendation is a typical application of recommender systems based on implicit feedbacks, where no explicit preferences (e.g., ratings) but only positive observations (e.g., clicks) are available \cite{mild2003improved,he2016fast,ren2017social}. These positive observations are usually in a form of sequential data as obtained by passively tracking users' behavior over a sequence of time. In this section, we briefly review the related work on session-based recommendation from the following two aspects, i.e., traditional methods and deep learning based methods.
  
\subsection{Traditional Methods}
  
Typically, there are two traditional modeling paradigms, i.e., general recommender and sequential recommender.
  
\textbf{General recommender} is mainly based on item-to-item recommendation approaches. In this setting, an item-to-item similarity matrix is pre-computed from the available session data. Items that are often clicked together (i.e., co-occurrence) in sessions are considered to be similar. \citet{linden2003amazon} propose an item-to-item collaborative filtering method to personalize the online store for each customer. \citet{sarwar2001item} analyze different item-based recommendation generation algorithms and compare their results with basic k-nearest neighbor approaches. Though these methods have proven to be effective and are widely employed, they only take into account the last click of the session, ignoring the information of the whole click sequence. 
  
\textbf{Sequential recommender} is based on Markov chains which utilizes sequential data by predicting users' next action given the last action~\cite{zimdars2001using,shani2005mdp}. \citet{zimdars2001using} propose a sequential recommender based on Markov chains and investigate how to extract sequential patterns to learn the next state using probabilistic decision-tree models. \citet{shani2005mdp} present a Markov Decesion Processes (MDP) aiming to provide recommendations in a session-based manner and the simplest MDP boil down to first-order Markov chains where the next recommendation can be simply computed through the transition probabilities between items. \citet{mobasher2002using} study different sequential patterns for recommendation and find that contiguous sequential patterns are more suitable for sequential prediction task than general sequential patterns. \citet{yap2012effective} introduce a new Competence Score measure in personalized sequential pattern mining for next-item recommendations. \citet{chen2012playlist} model playlists as Markov chains, and propose logistic Markov Embeddings to learn the representations of songs for playlists prediction. A major issue with applying Markov chains in the session-based recommendation task is that the state space quickly becomes unmanageable when trying to include all possible sequences of potential user selections over all items.
  
\subsection{Deep Learning based Methods}
  
Deep learning has recently been applied very successfully in areas such as image recognition \cite{krizhevsky2012imagenet,he2016deep}, speech recognition \cite{graves2013speech,amodei2016deep,hinton2012deep} and neural language processing \cite{socher2011parsing,de2014Medical,rsoy2014deep,song2017summarizing,li2017salience}. Deep models can be trained to learn discriminative representations from unstructured data \cite{he2017neural,he2017neuralfact,li2017neural}. Here, we focus on the related work that uses deep learning models to solve recommendation tasks.
  
\textbf{Neural network recommender} is mostly focusing on the classical collaborative filtering user-item setting. \citet{salakhutdinov2007restricted} first propose to use Restricted Boltzmann Machines (RBM) for Collaborative Filtering (CF). In their work, RBM is used to model user-item interactions and to perform recommendations. Recently, denoising auto-encoders have been used to perform CF in a similar manner \cite{wu2016collaborative,sedhain2015autorec}. \citet{wang2015learning} introduce a hierarchical representation model for the next basket recommendation which is based on encoder-decoder mechanism. Deep neural networks have also been used in cross-domain recommendations whereby items are mapped to a joint latent space \cite{elkahky2015a}. Recurrent Neural Networks (RNN) have been devised to model variable-length sequence data. Recently, \citet{hidasi2015session} apply RNN to session-based recommendation and achieve significant improvements over traditional methods. The proposed model utilizes session-parallel mini-batch training and employs ranking-based loss functions for learning the model. \citet{tan2016improved} further study the application of RNN in session-based recommendation. They propose two techniques to improve the performance of their model, namely data augmentation and a method to account for shifts in the input data distribution. \citet{zhang2014sequential} also use RNN for the click sequence prediction, they consider historical user behaviors as well as hand-crafted features for each user and item.
  
Though a growing number of publications on session-based recommendation focus on RNN-based methods, unlike existing studies, we propose a novel neural attentive recommendation model that combines both the user's sequential behavior and main purpose in the current session, which to the best of our knowledge, is not considered by existing researches. And we apply the attention mechanism to session-based recommendation for the first time.

\section{METHOD}

In this section, we first introduce the session-based recommendation task. Then we describe the proposed NARM in detail.
  
\subsection{Session-based Recommendation}
Session-based recommendation is the task of predicting what a user would like to click next when his/her current sequential transaction data is given. Here we give a formulation of the session-based recommendation problem.
  
Let $[x_{1},x_{2},...,x_{n-1},x_{n}]$ be a click session, where $x_{i}\in\mathcal{I} \,(1 \leq i \leq n)$ is the index of one clicked item out of a total number of $m$ items. We build a model $\mathbf{M}$ so that for any given prefix of the click sequence in the session, $\textbf{x} = [x_{1},x_{2},...,x_{t-1},x_{t}], 1 \leq t \leq n$, we get the output $\textbf{y} = \mathbf{M}(\textbf{x})$, where $\textbf{y} = [y_{1},y_{2},...,y_{m-1},y_{m}]$. We view $\textbf{y}$ as a ranking list over all the next items that can occur in that session, where $y_{j} \,(1 \leq j \leq m)$ corresponds to the recommendation score of item $j$. Since a recommender typically needs to make more than one recommendations for the user, thus the top-$k$ $(1 \leq k \leq m)$ items in $\textbf{y}$ are recommended. 
  
    \begin{figure}
  	    \vspace{1em}
	    \centering
	    \includegraphics[height=2.2in, width=2.2in]{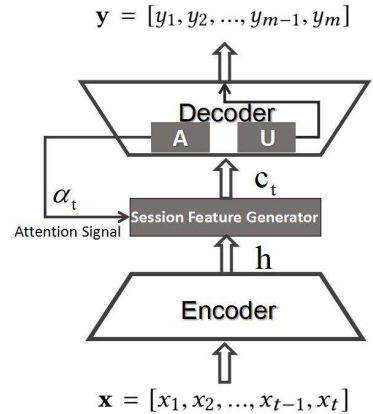}
	    \caption{The general framework and dataflow of the encoder-decoder-based NARM.}
    \end{figure}
  
    \begin{figure*}
	    \centering
        \subfloat[The graphical model of the global encoder in NARM, where the last hidden state is interpreted as the user's sequential behavior feature $\bm{c}_{t}^{\mathrm{g}}=\bm{h}_{t}$.]{
	        \includegraphics[height=1.5in, width=3.2in]{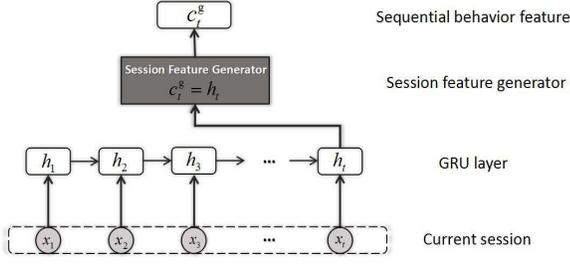}}
        \hspace{3.3em}
        \subfloat[The graphical model of the local encoder in NARM, where the weighted sum of hidden states is interpreted as the user's main purpose feature $\bm{c}_{t}^{\mathrm{l}} = \sum_{j=1}^{t} \alpha_{tj}\bm{h}_{j}$.]{
	        \includegraphics[height=1.55in, width=3.3in]{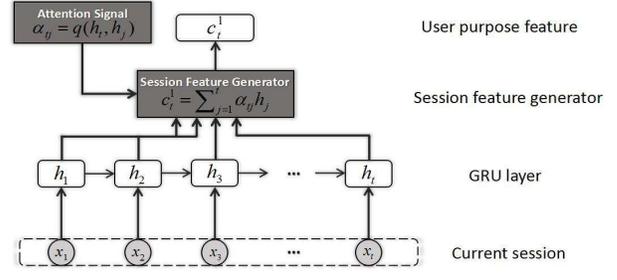}}	
        \caption{The global encoder and the local encoder in NARM.}
    \end{figure*}
         
\subsection{Overview}
In this paper, we propose an improved neural encoder-decoder architecture \cite{shang2015neural,ren2017leveraging} to address the session-based recommendation problem, named Neural Attentive Recommendation Machine (NARM). The basic idea of NARM is to build a hidden representation of the current session, and then generate predictions based on it. As shown in Figure 2, the encoder converts the input click sequence $\textbf{x} = [x_{1},x_{2},...,x_{t-1},{x_t}]$ into a set of high-dimensional hidden representations $\textbf{h} = [\textbf{\textsl{h}}_{1},\textbf{\textsl{h}}_{2},...,\textbf{\textsl{h}}_{t-1},\textbf{\textsl{h}}_{t}]$, which along with the attention signal at time $t$ (denoted as $\alpha_{t}$), are fed to the session feature generator to build the representation of the current session to decode at time $t$ (denoted as $\textbf{\textsl{c}}_{t}$). Finally $\textbf{\textsl{c}}_{t}$ is transformed by a matrix $\textbf{\textsl{U}}$ (as part of the decoder) into an activate function to produce a ranking list over all items, $\textbf{y} = [y_{1},y_{2},...,y_{m-1},y_{m}]$, that can occur in the current session.
    
The role of $\alpha_{t}$ is to determine which part of the hidden representations should be emphasized or ignored at time $t$. It should be noted that $\alpha_{t}$ could be fixed over time or changes dynamically during the prediction process. In the dynamic setting, $\alpha_{t}$ can be a function of the representations of hidden states or the input item embeddings. We adopt the dynamic setting in our model, more details will be described in \S 3.4.
  
The basic idea of our work is to learn a recommendation model that takes into consideration both the user's sequential behavior and main purpose in the current session. In the following part of this section, we first describe the global encoder in NARM which is used to model the user's sequential behavior (\S 3.3). Then we introduce the local encoder which is used to capture the user's main purpose in the current session (\S 3.4). Finally we show our NARM which combines both of them and computes the recommendation scores for each candidate item by using a bi-linear matching scheme (\S 3.5).
    
\subsection{Global Encoder in NARM}
In the global encoder, the inputs are entire previous clicks while the output is the feature of the user's sequential behavior in the current session. Both the inputs and output are uniformly represented by high-dimensional vectors.
  
Figure 3(a) shows the graphical model of the global encoder in NARM. We use a RNN with Gated Recurrent Units (GRU)  rather than a standard RNN because \citet{hidasi2015session} demonstrate that GRU can outperform the Long Short-Term Memory (LSTM) \cite{hochreiter2012long} units for the session-based recommendation task. GRU is a more elaborate RNN unit that aims at dealing with the vanishing gradient problem. The activation of GRU is a linear interpolation between the previous activation $\textbf{\textsl{h}}_{t-1}$ and the candidate activation $\widehat{\textbf{\textsl{h}}}_{t}$,
    \begin{equation}
  	    \textbf{\textsl{h}}_{t} = (1-\textbf{\textsl{z}}_{t})\textbf{\textsl{h}}_{t-1} + \textbf{\textsl{z}}_{t}\widehat{\textbf{\textsl{h}}}_{t} \;,
    \end{equation}
where the update gate $\bm{z_{t}}$ is given by
    \begin{equation}
  	    \textbf{\textsl{z}}_{t} = \sigma(\textbf{\textsl{W}}_{z}\textbf{\textsl{x}}_{t} + \textbf{\textsl{U}}_{z}\textbf{\textsl{h}}_{t-1})\;.
    \end{equation}
The candidate activation function $\widehat{\textbf{\textsl{h}}}_{t}$ is computed as
    \begin{equation}
    	\widehat{\textbf{\textsl{h}}}_{t} = tanh[\textbf{\textsl{W}}\textbf{\textsl{x}}_{t} + \textbf{\textsl{U}}(\textbf{\textsl{r}}_{t} \odot \textbf{\textsl{h}}_{t-1})] \;,
    \end{equation}
where the reset gate $\bm{r_{t}}$ is given by
    \begin{equation}
  	    \textbf{\textsl{r}}_{t} = \sigma(\textbf{\textsl{W}}_{r}\textbf{\textsl{x}}_{t} + \textbf{\textsl{U}}_{r}\textbf{\textsl{h}}_{t-1}) \;.
    \end{equation}
With a trivial session feature generator, we essentially use the final hidden state $\textbf{\textsl{h}}_{t}$ as the representation of the user's sequential behavior
    \begin{equation}
    	\textbf{\textsl{c}}_{t}^{\mathrm{g}} = \textbf{\textsl{h}}_{t} \;.
    \end{equation}
  
However, this global encoder has its drawbacks such as a vectorial summarization of the whole sequence behavior is often hard to capture a preciser intention of the current user.
  
\subsection{Local Encoder in NARM}

    \begin{figure*}
       	\centering
       	\includegraphics[height=2.3in, width=5.8in]{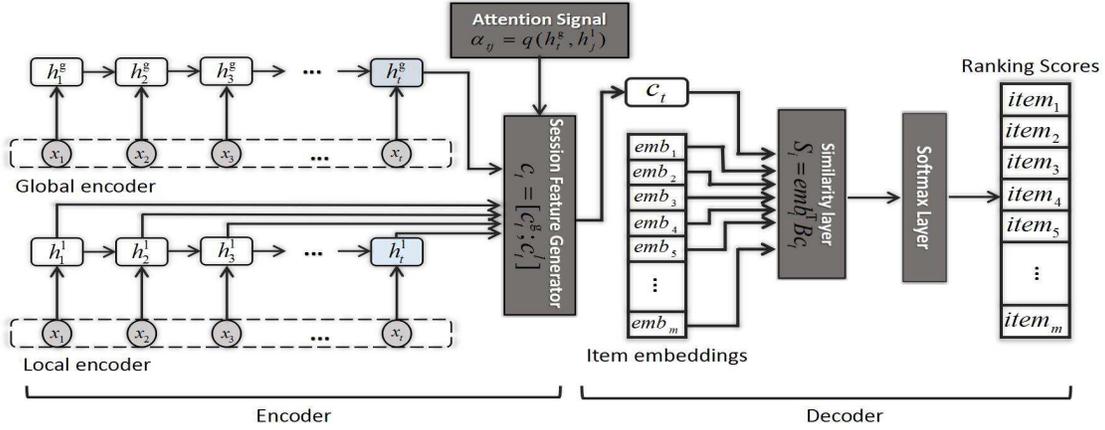}
       	\caption{The graphical model of NARM, where the session feature $\bm{c}_{t}$ is represented by the concatenation of vectors $\bm{c}_{t}^{\mathrm{g}}$ and $\bm{c}_{t}^{\mathrm{l}}$ (as computed in equation (5) and (6)). Note that $\bm{h}_{t}^{\mathrm{g}}$ and $\bm{h}_{t}^{\mathrm{l}}$ play different roles, while they have the same values. The last hidden state of the global encoder $\bm{h}_{t}^{\mathrm{g}}$ plays a role to encode the entire input clicks while the last hidden state of the local encoder $\bm{h}_{t}^{\mathrm{l}}$ is used to compute attention weights with the previous hidden states.}
    \end{figure*}

The architecture of the local encoder is similar to the global encoder as shown in Figure 3(b). In this encoding scheme we also use RNN with GRU as the basic component. To capture the user's main purpose in the current session, we involve an item-level attention mechanism which allows the decoder to dynamically select and linearly combine different parts of the input sequence, 
    \begin{equation}
  	    \bm{c}_{t}^{\mathrm{l}} = \sum_{j=1}^{t} \alpha_{tj}\textbf{\textsl{h}}_{j} \;,
    \end{equation} where the weighted factors $\alpha$ determine which part of the input sequence should be emphasized or ignored when making predictions, which in turn is a function of hidden states, 
    \begin{equation}
  	    \alpha_{tj} = q(\textbf{\textsl{h}}_{t},\textbf{\textsl{h}}_{j}) \;.
    \end{equation} 
Basically, the weighted factor $\alpha_{tj}$ models the alignment between the inputs around position $j$ and the output at position $t$, so it can be viewed as a specific matching model. In the local encoder, the function $q$ specifically computes the similarity between the final hidden state $\textbf{\textsl{h}}_{t}$ and the representation of the previous clicked item $\textbf{\textsl{h}}_{j}$,
    \begin{equation}
  	    q(\textbf{\textsl{h}}_{t},\textbf{\textsl{h}}_{j}) = \textbf{\textsl{v}}^{\mathrm{T}}\sigma(\textbf{\textsl{A}}_{1}\textbf{\textsl{h}}_{t} + \textbf{\textsl{A}}_{2}\textbf{\textsl{h}}_{j}) \;,
    \end{equation}
where $\sigma$ is an activate function such as sigmoid function, matrix $\textbf{\textsl{A}}_{1}$ is used to transform $\textbf{\textsl{h}}_{t}$ into a latent space, and $\textbf{\textsl{A}}_{2}$ plays the same role for $\textbf{\textsl{h}}_{j}$. 
  
This local encoder enjoys the advantages of adaptively focusing on more important items to capture the user's main purpose in the current session.

\subsection{NARM Model}
For the task of session-based recommendation, the global encoder has the summarization of the whole sequential behavior, while the local encoder can adaptively select the important items in the current session to capture the user's main purpose. We conjecture that the representation of the sequential behavior may provide useful information for capturing the user's main purpose in the current session. Therefore, we use the representations of the sequential behavior and the previous hidden states to compute the attention weight for each clicked item. Then a natural extension combines the sequential behavior feature and the user purpose feature by concatenating them to form an extended representation for each time stamp.
  
As shown in Figure 4, we can see the summarization $\bm{h}_{t}^{\mathrm{g}}$ is incorporated into $\bm{c}_{t}$ to provide a sequential behavior representation for NARM. It should be noticed that the session feature generator in NARM will evoke different encoding mechanisms in the global encoder and the local encoder, although they will be combined later to form a unified representation. More specifically, the last hidden state of the global encoder $\bm{h}_{t}^{\mathrm{g}}$ plays a role different from that of the local encoder $\bm{h}_{t}^{\mathrm{l}}$. The former has the responsibility to encode the entire sequential behavior. The latter is used to compute the attention weights with the previous hidden states. By this hybrid encoding scheme, both the user's sequential behavior and main purpose in the current session can be modeled into a unified representation $\bm{c}_{t}$, which is the concatenation of vectors $\bm{c}_{t}^{\mathrm{g}}$ and $\bm{c}_{t}^{\mathrm{l}}$,
    \begin{equation}
   	    \bm{c}_{t} = [\bm{c}_{t}^{\mathrm{g}};\bm{c}_{t}^{\mathrm{l}}] = [\bm{h}_{t}^{\mathrm{g}};\sum_{j=1}^{t} \alpha_{tj} \bm{h}_{t}^{\mathrm{l}}] \;.
    \end{equation}
  
    \begin{table*}
        \caption{Statistics of the datasets used in our experiments. (The avg.length means the average length of the complete dataset.)}
        \label{tab:freq}
            \begin{tabular}{lccccc}
            	\toprule
            	Datasets & all the clicks & train sessions & test sessions & all the items & avg.length\\
            	\midrule
            	YOOCHOOSE $1/64$ & 557248 & 369859 & 55898 & 16766 & 6.16\\
            	YOOCHOOSE $1/4$ & 8326407 & 5917746 & 55898 & 29618 & 5.71\\
            	DIGINETICA & 982961 & 719470 & 60858 & 43097 & 5.12\\
            	\bottomrule
            \end{tabular}
    \end{table*}
            
Figure 4 also gives a graphical illustration of the adopted decoding mechanism in NARM. Generally, a standard RNN utilizes fully-connected layer to decode. But using fully-connected layer means that the number of parameters to be learned in this layer is $|H| \ast |N|$ where $|H|$ is the dimension of the session representation and $|N|$ is the number of candidate items for prediction. Thus we have to reserve a large space to store these parameters. Though there are some approaches to reduce the parameters such as using a hierarchical softmax layer \cite{mnih2009scalable}, and negative sampling at random \cite{mikolov2013distributed}, they are not the best choices for our model.
  
We propose an alternative bi-linear decoding scheme which not only reduces the number of the parameters, but also improves the performance of NARM. Specifically, a bi-linear similarity function between the representations of the current session and each candidate items is used to compute a similarity score $S_{i}$,
    \begin{equation}
         S_{i} = {emb}_{i}^{\text{T}}\bm{B}\,\bm{c}_{t} \;,
    \end{equation}  
where $\textbf{\textsl{B}}$ is a $|D| \ast |H|$ matrix, $|D|$ is the dimension of each item embedding. Then the similarity score of each item is entered to a softmax layer to obtain the probability that the item will occur next. By using this bi-linear decoder, we reduce the number of parameters from $|N| \ast |H|$ to $|D| \ast |H|$, where $|D|$ is usually smaller than $|N|$. Moreover, the experiment results demonstrate that using this bi-linear decoder can improve the performance of NARM (as demonstrated in \S 4.4). 
  
To learn the parameters of the model, we do not utilize the proposed training procedure in \cite{hidasi2015session}, where the model is trained in a session-parallel, sequence-to-sequence manner. Instead, in order to fit the attention mechanism in the local encoder, NARM process each sequence $[x_{1},x_{2},...,x_{t-1},x_{t}]$ separately. Our model can be trained by using a standard mini-batch gradient descent on the cross-entropy loss:
    \begin{equation}
  	    L(p,q)=-\sum_{i=1}^{m}p_{i}log(q_{i})
    \end{equation}
where $q$ is the prediction probability distribution and $p$ is the truly distribution. At last, a Back-Propagation Through Time (BPTT) method for a fixed number of time steps is adopted to train NARM.

\section{EXPERIMENTAL SETUP}
In this section, we first describe the datasets, the state-of-the-art methods and the evaluation metrics employed in our experiments. Then we compare NARMs with different decoding schemes. Finally, we compare NARM with state-of-the-art methods.

\subsection{Dataset}
We evaluate different recommenders on two standard transaction datasets, i.e., YOOCHOOSE dataset and DIGINETICA dataset.

    \begin{itemize}
  	    \item YOOCHOOSE\footnote{http://2015.recsyschallenge.com/challenge.html} is a public dataset released by RecSys Challenge 2015. This dataset contains click-streams on an e-commerce site. After filtering out sessions of length 1 and items that appear less than 5 times, there remains 7981580 sessions and 37483 items.
  	    \item DIGINETICA\footnote{http://cikm2016.cs.iupui.edu/cikm-cup} comes from CIKM Cup 2016. We only used the released transaction data and also filtered out sessions of length 1 and items that appear less than 5 times. Finally the dataset contains 204771 sessions and 43097 items.
    \end{itemize}

We first conducted some preprocesses over two datasets. For YOOCHOOSE, we used the sessions of subsequent day for testing and filtered out clicks from the test set where the clicked items did not appear in the training set. For DIGINETICA, the only difference is that we use the sessions of subsequent week for testing. Because we did not train NARM in a session-parallel manner \cite{hidasi2015session}, a sequence splitting preprocess is necessary. For the input session $[x_{1},x_{2},...,x_{n-1},x_{n}]$, we generated the sequences and corresponding labels $([x_{1}],V(x_{2})$, $([x_{1},x_{2}],V(x_{3})$, ..., $([x_{1},x_{2},...,x_{n-1}],V(x_{n}))$ for training on both YOOCHOOSE and DIGINETICA. The corresponding label $V(x_{i})$ is the last click in the current session.
  
For the following reasons: (1) YOOCHOOSE is quite large, (2) \citet{tan2016improved} verified that the recommendation models do need to account for changing user behavior over time, (3) their experimental results showed that training on the entire dataset yields slightly poorer results than training on more recent fractions of the datasets. Thus we sorted the training sequences of YOOCHOOSE by time and reported our results on the model trained on more recent fractions $1/64$ and $1/4$ of training sequences as well. Note that some items that in the test set would not appear in the training set since we trained the model only on more recent fractions. The statistics of the three datasets (i.e., YOOCHOOSE $1/64$, YOOCHOOSE $1/4$ and DIGINETICA) are shown in Table 1.
    
    \begin{table*}
        \caption{The comparison of different decoders in NARM.}
	    \label{tab: decoder comparision}
	    \begin{tabular}{lcccccc}
		    \toprule
    	    \multirow{3}{*}{Decoders} & \multicolumn{2}{c}{YOOCHOOSE $1/64$} & \multicolumn{2}{c}{YOOCHOOSE $1/4$} & \multicolumn{2}{c}{DIGINETICA}\\
    	    \cmidrule(lr){2-3} \cmidrule(lr){4-5}\cmidrule(lr){6-7}
		        & Recall@20(\%) & MRR@20(\%) & Recall@20(\%) & MRR@20(\%) & Recall@20(\%) & MRR@20(\%)\\
    	    \midrule
    		    Fully-connected decoder & 67.67 & 29.17 & 69.49 & 29.54 & 57.84 & 24.77\\
    		    Bi-linear similarity decoder & \textbf{68.32} & 28.76 & \textbf{69.73} & 29.23 & \textbf{62.58} & \textbf{27.35}\\
    	    \bottomrule
	   \end{tabular}
    \end{table*}
    
    \begin{table*}
  	    \vspace{1.2em}
  	    \begin{threeparttable}
	        \caption{Performance comparison of NARM with baseline methods over three datasets.}
	        \label{tab:decoder comparision}
            \begin{tabular}{lcccccc}
		        \toprule
		        \multirow{3}{*}{Methods} & \multicolumn{2}{c}{YOOCHOOSE $1/64$} & \multicolumn{2}{c}{YOOCHOOSE $1/4$} & \multicolumn{2}{c}{DIGINETICA}\\
    	        \cmidrule(lr){2-3} \cmidrule(lr){4-5}\cmidrule(lr){6-7}
			        & Recall@20(\%) & MRR@20(\%) & Recall@20(\%) & MRR@20(\%) & Recall@20(\%) & MRR@20(\%)\\
    	        \midrule
    		        POP & 6.71 & 1.65 & 1.33 & 0.30 & 0.91 & 0.23\\
    		        S-POP & 30.44 & 18.35 & 27.08 & 17.75 & 21.07 & 14.69\\
    		        Item-KNN & 51.60 & 21.81 & 52.31 & 21.70 & 28.35 & 9.45\\
    		        BPR-MF & 31.31 & 12.08 & 3.40 & 1.57 & 15.19 & 8.63\\
    		        FPMC\tnote{*} & 45.62 & 15.01 & - & - & 31.55 & 8.92\\
    	        \midrule
    		        GRU-Rec & 60.64 & 22.89 & 59.53 & 22.60 & 43.82 & 15.46\\
    		        Improved GRU-Rec & 67.84 & \textbf{29.00} & 69.11 & 29.22 & 57.95 & 24.93\\
    		        NARM & \textbf{68.32} & 28.76 & \textbf{69.73} & \textbf{29.23} & \textbf{62.58} & \textbf{27.35}\\
    	        \bottomrule
            \end{tabular}
      
            \begin{tablenotes}
      	        \item[*] On YOOCHOOSE $1/4$, we do not have enough memory to initialize FPMC. Our available memory is 120G.
            \end{tablenotes}
      
        \end{threeparttable}
    \end{table*}

\subsection{Baseline Methods}
We compare the proposed NARM with five traditional methods (i.e., POP, S-POP, Item-KNN, BPR-MF and FPMC) and two RNN-based models (i.e., GRU-Rec and Improved GRU-Rec).
  
    \begin{itemize}
  	    \item \textbf{POP}: Popular predictor always recommends the most popular items in the training set. Despite its simplicity, it is often a strong baseline in certain domains.
  	    \item \textbf{S-POP}: This baseline recommends the most popular items for the current session. The recommendation list changes during the session gains more items. Ties are broken up using global popularity values.
  	    \item \textbf{Item-KNN}: In this baseline, similarity is defined as the co-occurrence number of two items in sessions divided by the square root of the product of the number of sessions in which either item occurs. Regularization is also included to avoid coincidental high similarities between rarely visited items \cite{linden2003amazon,davidson2010youtube}.
  	    \item \textbf{BPR-MF}: BPR-MF \cite{rendle2009bpr} optimizes a pairwise ranking objective function via stochastic gradient descent. Matrix factorization can not be directly applied to session-based recommendation because new sessions do not have precomputed latent representations. However, we can make it work by representing a new session with the average latent factors of items occurred in the session so far. In other words, the recommendation score can be computed as the average of the similarities between latent factors of a candidate item and the items in the session so far.
  	    \item \textbf{FPMC}: FPMC \cite{rendle2010factorizing} is a state-of-the-art hybrid model on the next-basket recommendation. In order to make it work on session-based recommendation, we do not consider the user latent representations when computing recommendation scores.
  	    \item \textbf{GRU-Rec}: We denote the model proposed in \cite{hidasi2015session} as GRU-Rec, which utilizes session-parallel mini-batch training process and also employs ranking-based loss functions for learning the model.
  	    \item \textbf{Improved GRU-Rec}: We denote the model proposed in \cite{tan2016improved} as Improved GRU-Rec. Improved GRU-Rec adopts two techniques which include data augmentation and a method to account for shifts in the input data distribution to improve the performance of GRU-Rec.
    \end{itemize}

\subsection{Evaluation Metrics and Experimental Setup}
  
\subsubsection{Evaluation Metrics}\
  
As recommender systems can only recommend a few items at each time, the actual item a user might pick should be amongst the first few items of the list. Therefore, we use the following metrics to evaluate the quality of the recommendation lists.

    \begin{itemize}
	    \item Recall@20: The primary evaluation metric is Recall@20 that is the proportion of cases when the desired item is amongst the top-20 items in all test cases. Recall@N does not consider the actual rank of the item as long as it is amongst the top-N and also usually correlates well with other metrics such as click-through rate (CTR) \cite{liu2012enlister}.
	    \item MRR@20: Another used metric is MRR@20 (Mean Reciprocal Rank), which is the average of reciprocal ranks of the desire items. The reciprocal rank is set to zero if the rank is larger than 20. MRR takes the rank of the item into account, which is important in settings where the order of recommendations matters.
    \end{itemize}
  
\subsubsection{Experimental Setup}\
  
The proposed NARM model uses 50-dimensional embeddings for the items. Optimization is done using Adam \cite{kingma2014adam} with the initial learning rate sets to 0.001, and the mini-batch size is fixed at 512. There are two dropout layers used in NARM: the first dropout layer is between the item embedding layer and the GRU layer with 25\% dropout, the second one is between the GRU layer and the bi-linear similarity layer with 50\% dropout. We also truncate BPTT at 19 time steps as the setting in the state-of-the-art method \cite{tan2016improved} and the number of epochs is set to 30 while using 10\% of the training data as the validation set. We use one GRU layer in our model and the GRU is set at 100 hidden units. The model is defined and trained in Theano on a GeForce GTX TitanX GPU. The source code of our model is available online\footnote{https://github.com/lijingsdu/sessionRec\_NARM}.

\subsection{Comparison among Different Decoders}
We first empirically compare NARMs with different decoders, i.e., fully-connected decoder and bi-linear similarity decoder. The results over three datasets are shown in Table 2. Here we only illustrate the results on 100-dimensional hidden states because we obtain the same conclusions on other dimension settings.
  
We make following observations from Table 2: (1) With regard to Recall@20, the performance improves when using the bi-linear similarity decoder, and the improvements are around 0.65\%, 0.24\% and 4.74\% respectively over three datasets. (2) And with regard to MRR@20, the performance on the model using the bi-linear decoder becomes a little worse on YOOCHOOSE $1/64$ and $1/4$. But on DIGINETICA, the model with the bi-linear decoder still obviously outperforms the model with the fully-connected decoder.
  
For the session-based recommendation task, as the recommender system recommends top-20 items at once in our settings, the actual item a user might pick should be among the list of 20 items. Thus we consider that the recall metric is more important than the MRR metric in this task, and NARM adopts the bi-linear decoder in the following experiments.

    \begin{figure*}[htbp]
	    \centering
	    \subfloat[YOOCHOOSE$1/64$]{
		    \includegraphics[height=3in, width=2in]{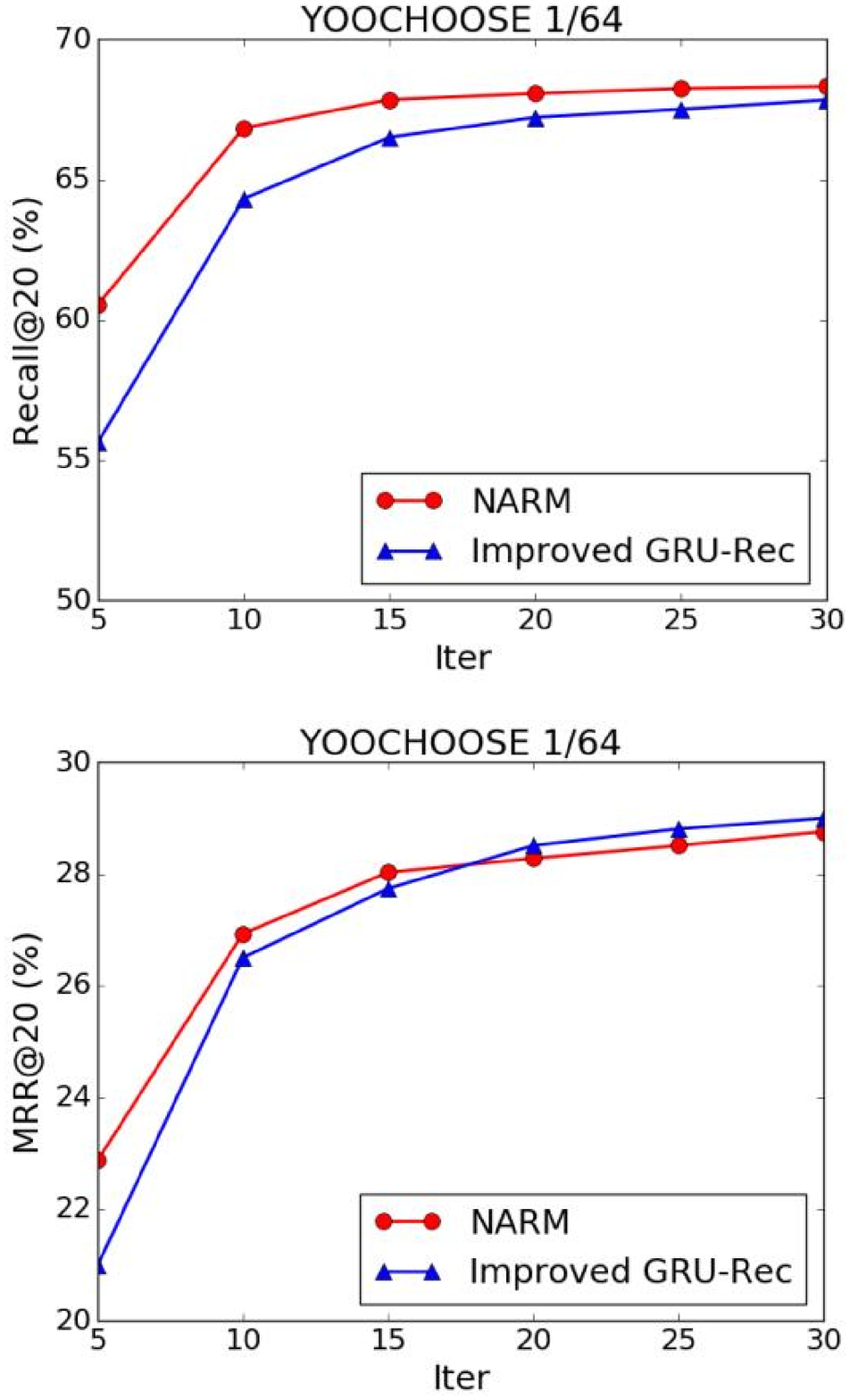}}
	    \vspace{0em} 
	    \subfloat[YOOCHOOSE$1/4$]{
		    \includegraphics[height=3in, width=2in]{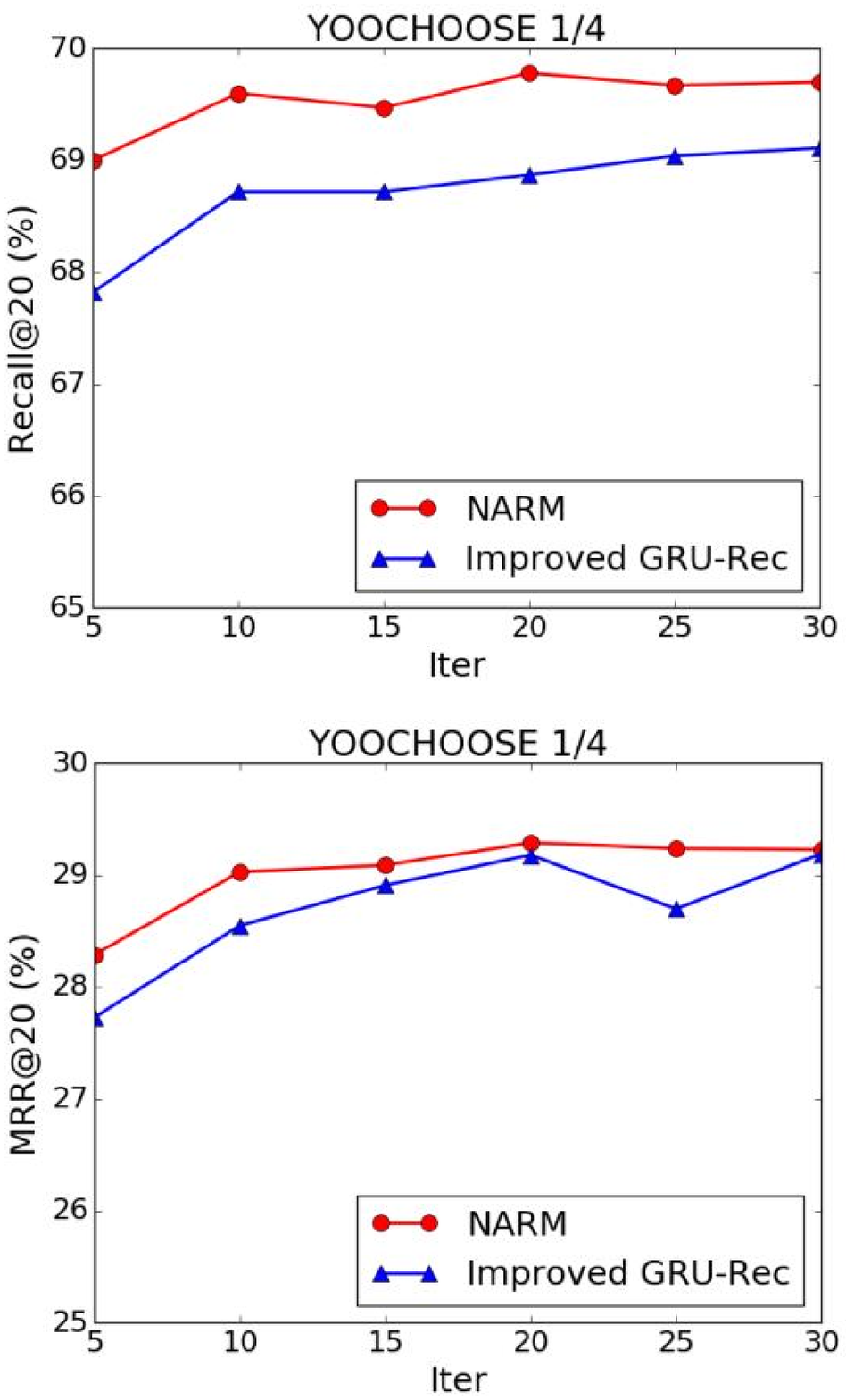}}	
	    \vspace{0em} 
	    \subfloat[DIGINETICA]{
		    \includegraphics[height=3in, width=2in]{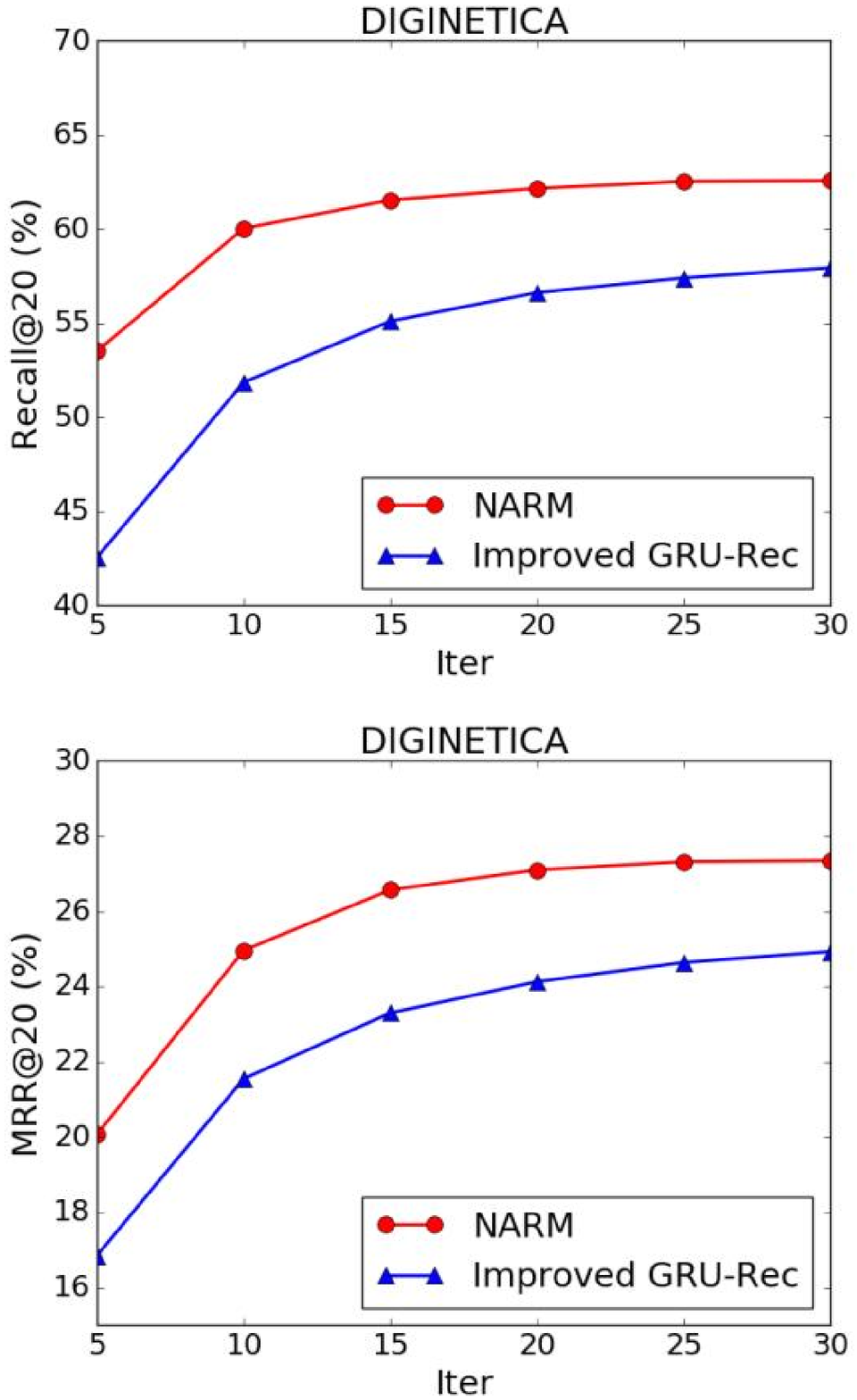}}
	    \caption{Performance comparison between NARM and the best baseline (i.e., Improved GRU-Rec) over three datasets.}
    \end{figure*}

\subsection{Comparison against Baselines}
Next we compare our NARM model with state-of-the-art methods. The results of all methods over three datasets are shown in Table 3. And a more specific comparison between NARM and the best baseline (i.e., Improved GRU-Rec) over three datasets are illustrated in Figure 5.

We have the following observations from the results: (1) For YOOCHOOSE $1/4$ dataset, BPR-MF does not work when we use the average of item factors occurred in the session to replace the user factor. Besides, since we regard each session as one user in FPMC, we do not have enough memory to initialize it. These problems indicate traditional user-based methods are no longer suitable for session-based recommendation. (2) Overall, three RNN-based methods consistently outperform the traditional baselines, which demonstrates that RNN-based models are good at dealing with sequence information in sessions. (3) By taking both the user's sequential behavior and main purpose into consideration, the proposed NARM can outperform all the baselines in terms of recall@20 over three datasets and can outperform most of the baselines in terms of MRR@20. Take DIGINETICA dataset as an example, when compared with the best baseline (i.e., Improved GRU-Rec), the relative performance improvements by NARM are around 7.98\% and 9.70\% respectively in terms of recall@20 and MRR@20. (4) As we can see, the recall values on two YOOCHOOSE datasets are not as significantly as the results on DIGINETICA and the obtained MRR values are very close to each other. We consider that one of the important reasons is when we split YOOCHOOSE dataset to $1/64$ and $1/4$, we do not filter out clicks from the test set where the clicked items are not in the training set in order to be consistent with the setting on Improved GRU-Rec \cite{tan2016improved}. While on DIGINETICA, we filter out these clicks from the test set, and hence NARM outperforms the baselines significantly in terms of both Recall@20 and MRR@20.

\section{ANALYSIS}
In this section, We further explore the influences of using different session features in NARM and analyze the effectiveness of the adopted attention mechanism.

    \begin{table}[htbp]
	    \vspace{-0.3em}
	    \caption{Performance comparison among three versions of NARM over three datasets.}
	    \centering
	    \subfloat[Performance comparison on YOOCHOOSE $1/64$]{
		    \begin{tabular}{lcccc}
			    \toprule
			    \multirow{3}{*}{Models} & \multicolumn{2}{c}{d=50} & \multicolumn{2}{c}{d=100}\\
			    & Recall@20 & MRR@20 & Recall@20 & MRR@20\\
			    \midrule
			    $NARM_{global}$ & 67.26 & 26.95 & 68.15 & 28.37\\
			    $NARM_{local}$ & 67.07 & 26.79 & 68.10 & 28.38\\
			    $NARM_{hybrid}$ & \textbf{68.28} & \textbf{28.10} & \textbf{68.32} & \textbf{28.76}\\
			    \bottomrule
		    \end{tabular}
	    }
	    
	    \subfloat[Performance comparison on YOOCHOOSE $1/4$]{
		    \begin{tabular}{lcccc}
			    \toprule
			    \multirow{3}{*}{Models} & \multicolumn{2}{c}{d=50} & \multicolumn{2}{c}{d=100}\\
			    & Recall@20 & MRR@20 & Recall@20 & MRR@20\\
			    \midrule
			    $NARM_{global}$ & 67.67 & 27.10 & 68.91 & 28.48\\
			    $NARM_{local}$ & 67.50 & 27.21 & 68.01 & 27.36\\
			    $NARM_{hybrid}$ & \textbf{69.17} & \textbf{28.67} & \textbf{69.73} & \textbf{29.23}\\
			    \bottomrule
		    \end{tabular}
    	}
 
	    \subfloat[Performance comparison on DIGINETICA]{
		    \begin{tabular}{lcccc}
			    \toprule
			    \multirow{3}{*}{Models} & \multicolumn{2}{c}{d=50} & \multicolumn{2}{c}{d=100}\\
			    & Recall@20 & MRR@20 & Recall@20 & MRR@20\\
			    \midrule
			    $NARM_{global}$ & 59.63 & 23.52 & 61.88 & 26.51\\
			    $NARM_{local}$ & 58.74 & 22.91 & 61.71 & 26.04\\
		    	$NARM_{hybrid}$ & \textbf{61.73} & \textbf{26.25} & \textbf{62.58} & \textbf{27.35}\\
			    \bottomrule
		    \end{tabular}
	    }
    \end{table}

\subsection{Influence of Using Different Features}
In this part, we refer to the NARM that uses the sequential behavior feature only, the NARM that uses the user purpose feature only, and the NARM that uses both two features as $NARM_{global}$, $NARM_{local}$ and $NARM_{hybrid}$ respectively. As shown in Table 4, (1) $NARM_{global}$ and $NARM_{local}$, which only use a single feature, do not perform well on three datasets. Besides, their performance are very close to each other in terms of two metrics. This indicates that merely considering the sequential behavior or the user purpose in the current session may not be able to learn a good recommendation model. (2) When we take into account both the user's sequential behavior and main purpose, $NARM_{hybrid}$ performs better than $NARM_{global}$ and $NARM_{local}$ in terms of Recall@20 and MRR@20 on different hidden state dimensions over three datasets. Take DIGINETICA dataset as an example, when compared with $NARM_{global}$ and $NARM_{local}$ with the dimensionality of the hidden state set to 50, the relative performance improvements by $NARM_{hybrid}$ are around 3.52\% and 5.09\% in terms of Recall@20 respectively. These results demonstrate the advantages of considering both the sequential behavior and the main purpose of the current user in session-based recommendation.
  
    \begin{figure*}
  	    \centering
  	    \includegraphics[height=1.7in, width=6.8in]{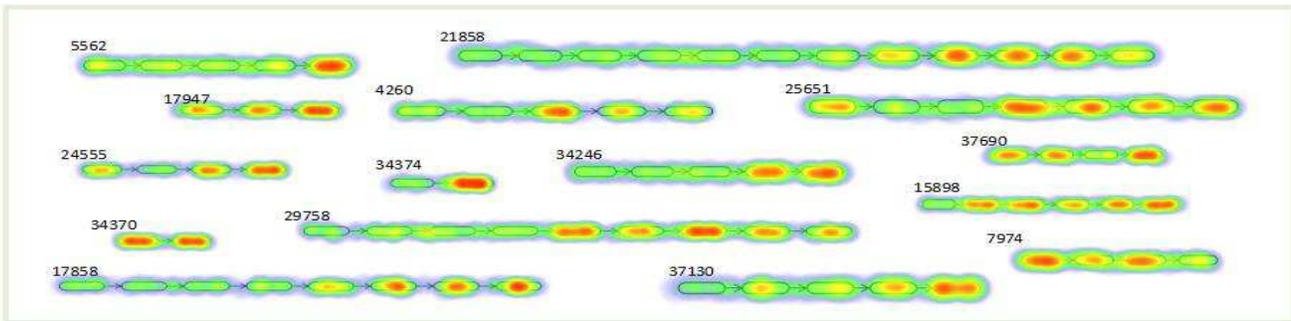}
  	    \caption{Visualization of items weights. The depth of the color corresponds to the importance of items given by equation (7). The numbers above the sessions is the session IDs. (Best viewed in color.)}
    \end{figure*}

    \begin{table}[htbp]
  	    \caption{Performance comparison among different session lengths on DIGINETICA dataset. (The baseline method is Improved GRU-Rec \cite{tan2016improved}.)}
  	    \label{tab:lenth comparision}
  	    \begin{tabular}{cccc}
  		    \toprule
  		    \multicolumn{4}{c}{DIGINETICA DATASET}\\
  		
  		    Length & Baseline correct & NARM correct & Performance\\
  		    \midrule
	  		1 & 8747 & 9358 & +6.98\%\\
	  		2 & 6601 & 7084 & +7.31\%\\
	  		3 & 4923 & 5299 & +7.63\%\\
	  		4 & 3625 & 3958 & +9.18\%\\
	  		5 & 2789 & 3019 & +8.24\%\\
	  		6 & 2029 & 2202 & +8.52\%\\
	  		7 & 1520 & 1656 & +8.94\%\\
	  		8 & 1198 & 1295 & +8.09\%\\
	  		9 & 915 & 996 & +8.85\%\\
	  		10 & 690 & 753 & +9.13\%\\
	  		11 & 509 & 587 & \textbf{+15.32\%}\\
	  		12 & 411 & 459 & \textbf{+11.67\%}\\
	  		13 & 304 & 323 & +6.25\%\\
	  		14 & 243 & 260 & +6.99\%\\
	  		15 & 199 & 219 & \textbf{+10.05\%}\\
	  		16 & 149 & 165 & \textbf{+10.73\%}\\
	  		17 & 98 & 112 &\textbf{+14.28\%}\\
	  		18 & 88 & 93 & +5.68\%\\
	  		19 & 70 & 75 & +7.14\%\\
  		    \bottomrule
  	    \end{tabular}
    \end{table}

\subsection{Influence of Different Session Lengths}
Our NARM model is based on the assumption that when a user is browsing online, his/her click behavior frequently revolves his/her main purpose in the current session. However, we can hardly capture the user's main purpose when s/he just clicks a few items. Therefore, our NARM model should be good at modeling long sessions. To verify this, we make comparisons among sessions with different lengths on DIGINETICA. As shown in Table 5, (1) NARM performs better when the session lengths are between 4 and 17 in general. This indicates that NARM do capture the user's main purpose more accuracy on long sessions. In other words, it could make a better prediction if NARM captures more user purpose features on the basis of the existing sequential behavior features. (2) When sessions are too long, the performance improvements of NARM are declined. We consider the reason is that when a session is too long, the user is very likely to click some items aimlessly, so that the local encoder in NARM could not capture the user's main purpose in the current session.

\subsection{Visualize the Attention Weights}
To illustrate the role of the attention mechanism intuitively, we present an example in Figure 6. The session instances are chosen randomly from DIGINETICA. The depth of the color corresponds to the importance of items given by equation (7). We have following observations from the example: (1) Overall, it is obvious that not all items are related to the next click and almost all the important items in the current session is continuous. This implies that the users' intentions in sessions are indeed localized, which is one of the reasons why NARM can outperform the general RNN-based model. (2) The most important items are often near the end of the session. This is in line with people's browsing behavior: a user is very likely to click other items that are related to what s/he has clicked just now. Recall that general RNN-based models are able to model this fact, thus they can achieve fairly good performance in session-based recommendation. (3) In some cases, the most important items appear in the beginning or middle of the session (e.g., in session 7974 or 4260). In this situation, we believe that our NARM can perform better than general RNN-based models because the attention mechanism could learn to pay more attention to more important items regardless of its position in one session. 

\section{CONCLUSION \& FUTURE WORK}
We have proposed the neural attentive recommendation machine (NARM) with an encoder-decoder architecture to address the session-based recommendation problem. By incorporating an attention mechanism into RNN, our proposed approach can capture both the user's sequential behavior and main purpose in the current session. \if0With this attention mechanism, NARM can attend differentially to more and less important items.\fi Based on the sequential behavior feature and the user purpose feature, we have applied NARM to predict a user's next click in the current session. We have conducted extensive experiments on two benchmark datasets and demonstrated that our approach can outperform state-of-the-art methods in terms of different evaluation metrics. Moreover, we have performed an analysis on user click behaviors and found that users' intentions are localized in most sessions, which proves the rationality of our model.

As to future work, more item attributes, such as prices and categories, may enhance the performance of our method in session-based recommendation. Meanwhile, both the nearest neighbor sessions and the importance of different neighbors  should give new insights. Finally, the attention mechanism can be used to explore the importance of attributes in the current session.

\section*{Acknowledgments}
The authors wish to thank the anonymous reviewers for their helpful comments. This work is supported by the Natural Science Foundation of China (61672322, 61672324), the Natural Science Foundation of Shandong province (2016ZRE27468) and the Fundamental Research Funds of Shandong University.

\bibliographystyle{abbrvnatnourl}
\bibliography{sigproc} 

\begin{thebibliography}{46}
\providecommand{\natexlab}[1]{#1}
\providecommand{\url}[1]{\texttt{#1}}
\expandafter\ifx\csname urlstyle\endcsname\relax
  \providecommand{\doi}[1]{doi: #1}\else
  \providecommand{\doi}{doi: \begingroup \urlstyle{rm}\Url}\fi

\bibitem[Adomavicius and Tuzhilin(2005)]{adomavicius2005toward}
G.~Adomavicius and A.~Tuzhilin.
\newblock Toward the next generation of recommender systems: a survey of the
  state-of-the-art and possible extensions.
\newblock \emph{IEEE Transactions on Knowledge and Data Engineering},
  17\penalty0 (6):\penalty0 734--749, 2005.

\bibitem[Amodei et~al.(2016)Amodei, Anubhai, Battenberg, Case, Casper,
  Catanzaro, Chen, Chrzanowski, Coates, Diamos, et~al.]{amodei2016deep}
D.~Amodei, R.~Anubhai, E.~Battenberg, C.~Case, J.~Casper, B.~Catanzaro,
  J.~Chen, M.~Chrzanowski, A.~Coates, G.~Diamos, et~al.
\newblock Deep speech 2: end-to-end speech recognition in english and mandarin.
\newblock In \emph{Proceedings of the 33rd. International Conference on Machine
  Learning}, pages 173--182, 2016.

\bibitem[Chen et~al.(2012)Chen, Moore, Turnbull, and
  Joachims]{chen2012playlist}
S.~Chen, J.~L. Moore, D.~Turnbull, and T.~Joachims.
\newblock Playlist prediction via metric embedding.
\newblock In \emph{Proceedings of the 18th. ACM SIGKDD International Conference
  on Knowledge Discovery and Data Mining}, pages 714--722, 2012.

\bibitem[Davidson et~al.(2010)Davidson, Liebald, Liu, Nandy, Van~Vleet, Gargi,
  Gupta, He, Lambert, Livingston, et~al.]{davidson2010youtube}
J.~Davidson, B.~Liebald, J.~Liu, P.~Nandy, T.~Van~Vleet, U.~Gargi, S.~Gupta,
  Y.~He, M.~Lambert, B.~Livingston, et~al.
\newblock The youtube video recommendation system.
\newblock In \emph{Proceedings of the 4th. ACM Conference on Recommender
  Systems}, pages 293--296, 2010.

\bibitem[De~Vine et~al.(2014)De~Vine, Zuccon, Koopman, Sitbon, and
  Bruza]{de2014Medical}
L.~De~Vine, G.~Zuccon, B.~Koopman, L.~Sitbon, and P.~Bruza.
\newblock Medical semantic similarity with a neural language model.
\newblock In \emph{Proceedings of the 23rd. ACM International Conference on
  Conference on Information and Knowledge Management}, pages 1819--1822, 2014.

\bibitem[Elkahky et~al.(2015)Elkahky, Song, and He]{elkahky2015a}
A.~M. Elkahky, Y.~Song, and X.~He.
\newblock A multi-view deep learning approach for cross domain user modeling in
  recommendation systems.
\newblock In \emph{Proceedings of the 24th. International Conference on World
  Wide Web}, pages 278--288, 2015.

\bibitem[Graves et~al.(2013)Graves, Mohamed, and Hinton]{graves2013speech}
A.~Graves, A.-r. Mohamed, and G.~Hinton.
\newblock Speech recognition with deep recurrent neural networks.
\newblock In \emph{Proceedings of the IEEE International Conference on
  Acoustics, Speech and Signal Processing}, pages 6645--6649, 2013.

\bibitem[He et~al.(2016{\natexlab{a}})He, Zhang, Ren, and Sun]{he2016deep}
K.~He, X.~Zhang, S.~Ren, and J.~Sun.
\newblock Deep residual learning for image recognition.
\newblock In \emph{Proceedings of the IEEE Conference on Computer Vision and
  Pattern Recognition}, pages 770--778, 2016{\natexlab{a}}.

\bibitem[He and Chua(2017)]{he2017neuralfact}
X.~He and T.-S. Chua.
\newblock Neural factorization machines for sparse predictive analytics.
\newblock In \emph{Proceedings of the 40th. International ACM SIGIR conference
  on Research and Development in Information Retrieval}, pages 355--364, 2017.

\bibitem[He et~al.(2016{\natexlab{b}})He, Zhang, Kan, and Chua]{he2016fast}
X.~He, H.~Zhang, M.-Y. Kan, and T.-S. Chua.
\newblock Fast matrix factorization for online recommendation with implicit
  feedback.
\newblock In \emph{Proceedings of the 39th. International ACM SIGIR conference
  on Research and Development in Information Retrieval}, pages 549--558,
  2016{\natexlab{b}}.

\bibitem[He et~al.(2017)He, Liao, Zhang, Nie, Hu, and Chua]{he2017neural}
X.~He, L.~Liao, H.~Zhang, L.~Nie, X.~Hu, and T.-S. Chua.
\newblock Neural collaborative filtering.
\newblock In \emph{Proceedings of the 26th. International Conference on World
  Wide Web}, pages 173--182, 2017.

\bibitem[Hidasi et~al.(2016)Hidasi, Karatzoglou, Baltrunas, and
  Tikk]{hidasi2015session}
B.~Hidasi, A.~Karatzoglou, L.~Baltrunas, and D.~Tikk.
\newblock Session-based recommendations with recurrent neural networks.
\newblock In \emph{Proceedings of the 4th. International Conference on Learning
  Representations}, 2016.

\bibitem[Hinton et~al.(2012)Hinton, Deng, Yu, Dahl, Mohamed, Jaitly, Senior,
  Vanhoucke, Nguyen, and Sainath]{hinton2012deep}
G.~Hinton, L.~Deng, D.~Yu, G.~E. Dahl, A.~R. Mohamed, N.~Jaitly, A.~Senior,
  V.~Vanhoucke, P.~Nguyen, and T.~N. Sainath.
\newblock Deep neural networks for acoustic modeling in speech recognition: the
  shared views of four research groups.
\newblock \emph{IEEE Signal Processing Magazine}, 29\penalty0 (6):\penalty0
  82--97, 2012.

\bibitem[Hochreiter and Schmidhuber(1997)]{hochreiter2012long}
S.~Hochreiter and J.~Schmidhuber.
\newblock Long short-term memory.
\newblock \emph{Neural computation}, 9\penalty0 (8):\penalty0 1735--1780, 1997.

\bibitem[Kingma and Ba(2015)]{kingma2014adam}
D.~Kingma and J.~Ba.
\newblock Adam: a method for stochastic optimization.
\newblock In \emph{Proceedings of the 4th. International Conference on Learning
  Representations}, 2015.

\bibitem[Koren et~al.(2009)Koren, Bell, and Volinsky]{koren2009matrix}
Y.~Koren, R.~Bell, and C.~Volinsky.
\newblock Matrix factorization techniques for recommender systems.
\newblock \emph{Computer}, 42\penalty0 (8):\penalty0 30--37, 2009.

\bibitem[Krizhevsky et~al.(2012)Krizhevsky, Sutskever, and
  Hinton]{krizhevsky2012imagenet}
A.~Krizhevsky, I.~Sutskever, and G.~E. Hinton.
\newblock Imagenet classification with deep convolutional neural networks.
\newblock In \emph{Proceedings of the 25th. International Conference on Neural
  Information Processing Systems}, pages 1097--1105, 2012.

\bibitem[Li et~al.(2017{\natexlab{a}})Li, Wang, Lam, Ren, and
  Bing]{li2017salience}
P.~Li, Z.~Wang, W.~Lam, Z.~Ren, and L.~Bing.
\newblock Salience estimation via variational auto-encoders for multi-document
  summarization.
\newblock In \emph{Proceedings of the 31st. AAAI Conference on Artificial
  Intelligence}, pages 3497--3503, 2017{\natexlab{a}}.

\bibitem[Li et~al.(2017{\natexlab{b}})Li, Wang, Ren, Bing, and
  Lam]{li2017neural}
P.~Li, Z.~Wang, Z.~Ren, L.~Bing, and W.~Lam.
\newblock Neural rating regression with abstractive tips generation for
  recommendation.
\newblock In \emph{Proceedings of the 40th. International ACM SIGIR conference
  on Research and Development in Information Retrieval}, pages 345--354,
  2017{\natexlab{b}}.

\bibitem[Linden et~al.(2003)Linden, Smith, and York]{linden2003amazon}
G.~Linden, B.~Smith, and J.~York.
\newblock Amazon.com recommendations: item-to-item collaborative filtering.
\newblock \emph{IEEE Internet Computing}, 7\penalty0 (1):\penalty0 76--80,
  2003.

\bibitem[Liu et~al.(2012)Liu, Chen, Cai, and Yu]{liu2012enlister}
Q.~Liu, T.~Chen, J.~Cai, and D.~Yu.
\newblock Enlister: baidu's recommender system for the biggest chinese q\&a
  website.
\newblock In \emph{Proceedings of the 6th. ACM Conference on Recommender
  Systems}, pages 285--288, 2012.

\bibitem[Mikolov et~al.(2013)Mikolov, Sutskever, Chen, Corrado, and
  Dean]{mikolov2013distributed}
T.~Mikolov, I.~Sutskever, K.~Chen, G.~Corrado, and J.~Dean.
\newblock Distributed representations of words and phrases and their
  compositionality.
\newblock In \emph{Proceedings of the 26th. International Conference on Neural
  Information Processing Systems}, pages 3111--3119, 2013.

\bibitem[Mild and Reutterer(2003)]{mild2003improved}
A.~Mild and T.~Reutterer.
\newblock An improved collaborative filtering approach for predicting
  cross-category purchases based on binary market basket data.
\newblock \emph{Journal of Retailing and Consumer Services}, 10\penalty0
  (3):\penalty0 123--133, 2003.

\bibitem[Mnih and Hinton(2008)]{mnih2009scalable}
A.~Mnih and G.~Hinton.
\newblock A scalable hierarchical distributed language model.
\newblock In \emph{Proceedings of the 21st. International Conference on Neural
  Information Processing Systems}, pages 1081--1088, 2008.

\bibitem[Mobasher et~al.(2002)Mobasher, Dai, Luo, and
  Nakagawa]{mobasher2002using}
B.~Mobasher, H.~Dai, T.~Luo, and M.~Nakagawa.
\newblock Using sequential and non-sequential patterns in predictive web usage
  mining tasks.
\newblock In \emph{Proceedings of the IEEE International Conference on Data
  Mining}, pages 669--672, 2002.

\bibitem[Ren et~al.(2017{\natexlab{a}})Ren, Chen, Ren, Wei, Ma, and
  de~Rijke]{ren2017leveraging}
P.~Ren, Z.~Chen, Z.~Ren, F.~Wei, J.~Ma, and M.~de~Rijke.
\newblock Leveraging contextual sentence relations for extractive summarization
  using a neural attention model.
\newblock In \emph{Proceedings of the 40th. International ACM SIGIR conference
  on Research and Development in Information Retrieval}, pages 95--104,
  2017{\natexlab{a}}.

\bibitem[Ren et~al.(2017{\natexlab{b}})Ren, Liang, Li, Wang, and
  de~Rijke]{ren2017social}
Z.~Ren, S.~Liang, P.~Li, S.~Wang, and M.~de~Rijke.
\newblock Social collaborative viewpoint regression with explainable
  recommendations.
\newblock In \emph{Proceedings of the 10th. ACM International Conference on Web
  Search and Data Mining}, pages 485--494, 2017{\natexlab{b}}.

\bibitem[Rendle et~al.(2009)Rendle, Freudenthaler, Gantner, and
  Schmidt-Thieme]{rendle2009bpr}
S.~Rendle, C.~Freudenthaler, Z.~Gantner, and L.~Schmidt-Thieme.
\newblock Bpr: bayesian personalized ranking from implicit feedback.
\newblock In \emph{Proceedings of the 25th. Conference on Uncertainty in
  Artificial Intelligence}, pages 452--461, 2009.

\bibitem[Rendle et~al.(2010)Rendle, Freudenthaler, and
  Schmidt-Thieme]{rendle2010factorizing}
S.~Rendle, C.~Freudenthaler, and L.~Schmidt-Thieme.
\newblock Factorizing personalized markov chains for next-basket
  recommendation.
\newblock In \emph{Proceedings of the 19th. International Conference on World
  Wide Web}, pages 811--820, 2010.

\bibitem[Rsoy and Cardie(2014)]{rsoy2014deep}
O.~Rsoy and C.~Cardie.
\newblock Deep recursive neural networks for compositionality in language.
\newblock In \emph{Proceedings of the 27th. International Conference on Neural
  Information Processing Systems}, pages 2096--2104, 2014.

\bibitem[Salakhutdinov et~al.(2007)Salakhutdinov, Mnih, and
  Hinton]{salakhutdinov2007restricted}
R.~Salakhutdinov, A.~Mnih, and G.~Hinton.
\newblock Restricted boltzmann machines for collaborative filtering.
\newblock In \emph{Proceedings of the 24th. International Conference on Machine
  Learning}, pages 791--798, 2007.

\bibitem[Sarwar et~al.(2001)Sarwar, Karypis, Konstan, and
  Riedl]{sarwar2001item}
B.~Sarwar, G.~Karypis, J.~Konstan, and J.~Riedl.
\newblock Item-based collaborative filtering recommendation algorithms.
\newblock In \emph{Proceedings of the 10th. International Conference on World
  Wide Web}, pages 285--295, 2001.

\bibitem[Schafer et~al.(1999)Schafer, Konstan, and
  Riedl]{schafer1999recommender}
J.~B. Schafer, J.~Konstan, and J.~Riedl.
\newblock Recommender systems in e-commerce.
\newblock In \emph{Proceedings of the 1st. ACM Conference on Electronic
  Commerce}, pages 158--166, 1999.

\bibitem[Sedhain et~al.(2015)Sedhain, Menon, Sanner, and
  Xie]{sedhain2015autorec}
S.~Sedhain, A.~K. Menon, S.~Sanner, and L.~Xie.
\newblock Autorec: autoencoders meet collaborative filtering.
\newblock In \emph{Proceedings of the 24th. International Conference on World
  Wide Web}, pages 111--112, 2015.

\bibitem[Shang et~al.(2015)Shang, Lu, and Li]{shang2015neural}
L.~Shang, Z.~Lu, and H.~Li.
\newblock Neural responding machine for short-text conversation.
\newblock In \emph{Proceedings of the 53rd. Annual Meeting of the Association
  for Computational Linguistics}, pages 1577--1586, 2015.

\bibitem[Shani et~al.(2005)Shani, Heckerman, and Brafman]{shani2005mdp}
G.~Shani, D.~Heckerman, and R.~I. Brafman.
\newblock An mdp-based recommender system.
\newblock \emph{Journal of Machine Learning Research}, 6\penalty0 (1):\penalty0
  1265--1295, 2005.

\bibitem[Socher et~al.(2011)Socher, Lin, Ng, and Manning]{socher2011parsing}
R.~Socher, C.~Y. Lin, A.~Y. Ng, and C.~D. Manning.
\newblock Parsing natural scenes and natural language with recursive neural
  networks.
\newblock In \emph{Proceedings of the 28th. International Conference on Machine
  Learning}, pages 129--136, 2011.

\bibitem[Song et~al.(2017)Song, Ren, Liang, Li, Ma, and
  de~Rijke]{song2017summarizing}
H.~Song, Z.~Ren, S.~Liang, P.~Li, J.~Ma, and M.~de~Rijke.
\newblock Summarizing answers in non-factoid community question-answering.
\newblock In \emph{Proceedings of the 10th. ACM International Conference on Web
  Search and Data Mining}, pages 405--414, 2017.

\bibitem[Su and Khoshgoftaar(2009)]{su2009survey}
X.~Su and T.~M. Khoshgoftaar.
\newblock A survey of collaborative filtering techniques.
\newblock \emph{Advances in Artificial Intelligence}, 2009.

\bibitem[Tan et~al.(2016)Tan, Xu, and Liu]{tan2016improved}
Y.~K. Tan, X.~Xu, and Y.~Liu.
\newblock Improved recurrent neural networks for session-based recommendations.
\newblock In \emph{Proceedings of the 1st. Workshop on Deep Learning for
  Recommender Systems}, pages 17--22, 2016.

\bibitem[Wang et~al.(2015)Wang, Guo, Lan, Xu, Wan, and Cheng]{wang2015learning}
P.~Wang, J.~Guo, Y.~Lan, J.~Xu, S.~Wan, and X.~Cheng.
\newblock Learning hierarchical representation model for nextbasket
  recommendation.
\newblock In \emph{Proceedings of the 38th. International ACM SIGIR conference
  on Research and Development in Information Retrieval}, pages 403--412, 2015.

\bibitem[Weimer et~al.(2007)Weimer, Karatzoglou, Le, and
  Smola]{weimer2007maximum}
M.~Weimer, A.~Karatzoglou, Q.~V. Le, and A.~Smola.
\newblock Maximum margin matrix factorization for collaborative ranking.
\newblock In \emph{Proceedings of the 20th. International Conference on Neural
  Information Processing Systems}, pages 1--8, 2007.

\bibitem[Wu et~al.(2016)Wu, Dubois, Zheng, and Ester]{wu2016collaborative}
Y.~Wu, C.~Dubois, A.~X. Zheng, and M.~Ester.
\newblock Collaborative denoising auto-encoders for top-n recommender systems.
\newblock In \emph{Proceedings of the 9th. ACM International Conference on Web
  Search and Data Mining}, pages 153--162, 2016.

\bibitem[Yap et~al.(2012)Yap, Li, and Yu]{yap2012effective}
G.~E. Yap, X.~L. Li, and P.~S. Yu.
\newblock Effective next-items recommendation via personalized sequential
  pattern mining.
\newblock In \emph{Proceedings of the 17th. International Conference on
  Database Systems for Advanced Applications}, pages 48--64, 2012.

\bibitem[Zhang et~al.(2014)Zhang, Dai, Xu, Feng, Wang, Bian, Wang, and
  Liu]{zhang2014sequential}
Y.~Zhang, H.~Dai, C.~Xu, J.~Feng, T.~Wang, J.~Bian, B.~Wang, and T.-Y. Liu.
\newblock Sequential click prediction for sponsored search with recurrent
  neural networks.
\newblock In \emph{Proceedings of the 28th. AAAI Conference on Artificial
  Intelligence}, pages 1369--1375, 2014.

\bibitem[Zimdars et~al.(2001)Zimdars, Chickering, and Meek]{zimdars2001using}
A.~Zimdars, D.~M. Chickering, and C.~Meek.
\newblock Using temporal data for making recommendations.
\newblock In \emph{Proceedings of the 17th. Conference on Uncertainty in
  Artificial Intelligence}, pages 580--588, 2001.

\end{thebibliography}



\begin{thebibliography}{00}

%%% the Web address.

\ifx \showCODEN    \undefined \def \showCODEN     #1{\unskip}     \fi
\ifx \showDOI      \undefined \def \showDOI       #1{{\tt DOI:}\penalty0{#1}\ }
  \fi
\ifx \showISBNx    \undefined \def \showISBNx     #1{\unskip}     \fi
\ifx \showISBNxiii \undefined \def \showISBNxiii  #1{\unskip}     \fi
\ifx \showISSN     \undefined \def \showISSN      #1{\unskip}     \fi
\ifx \showLCCN     \undefined \def \showLCCN      #1{\unskip}     \fi
\ifx \shownote     \undefined \def \shownote      #1{#1}          \fi
\ifx \showarticletitle \undefined \def \showarticletitle #1{#1}   \fi
\ifx \showURL      \undefined \def \showURL       {\relax}        \fi
% invisible to TeX
\providecommand\bibfield[2]{#2}
\providecommand\bibinfo[2]{#2}
\providecommand\natexlab[1]{#1}
\providecommand\showeprint[2][]{arXiv:#2}

\bibitem[\protect\citeauthoryear{Adomavicius and Tuzhilin}{Adomavicius and
  Tuzhilin}{2005}]%
        {adomavicius2005toward}
\bibfield{author}{\bibinfo{person}{Gediminas Adomavicius} {and}
  \bibinfo{person}{Alexander Tuzhilin}.} \bibinfo{year}{2005}\natexlab{}.
\newblock \showarticletitle{Toward the next generation of recommender systems:
  A survey of the state-of-the-art and possible extensions}.
\newblock \bibinfo{journal}{{\em IEEE Transactions on Knowledge and Data
  Engineering\/}} \bibinfo{volume}{17}, \bibinfo{number}{6}
  (\bibinfo{year}{2005}), \bibinfo{pages}{734--749}.
\newblock


\bibitem[\protect\citeauthoryear{Al-Rfou, Alain, Almahairi, Angermueller,
  Bahdanau, Ballas, Bastien, Bayer, Belikov, Belopolsky, et~al\mbox{.}}{Al-Rfou
  et~al\mbox{.}}{2016}]%
        {al2016theano}
\bibfield{author}{\bibinfo{person}{Rami Al-Rfou}, \bibinfo{person}{Guillaume
  Alain}, \bibinfo{person}{Amjad Almahairi}, \bibinfo{person}{Christof
  Angermueller}, \bibinfo{person}{Dzmitry Bahdanau}, \bibinfo{person}{Nicolas
  Ballas}, \bibinfo{person}{Fr{\'e}d{\'e}ric Bastien}, \bibinfo{person}{Justin
  Bayer}, \bibinfo{person}{Anatoly Belikov}, \bibinfo{person}{Alexander
  Belopolsky}, {and} \bibinfo{person}{others}.}
  \bibinfo{year}{2016}\natexlab{}.
\newblock \showarticletitle{Theano: A Python framework for fast computation of
  mathematical expressions}.
\newblock \bibinfo{journal}{{\em arXiv preprint arXiv:1605.02688\/}}
  (\bibinfo{year}{2016}).
\newblock


\bibitem[\protect\citeauthoryear{Amodei, Anubhai, Battenberg, Case, Casper,
  Catanzaro, Chen, Chrzanowski, Coates, Diamos, et~al\mbox{.}}{Amodei
  et~al\mbox{.}}{2016}]%
        {amodei2016deep}
\bibfield{author}{\bibinfo{person}{Dario Amodei}, \bibinfo{person}{Rishita
  Anubhai}, \bibinfo{person}{Eric Battenberg}, \bibinfo{person}{Carl Case},
  \bibinfo{person}{Jared Casper}, \bibinfo{person}{Bryan Catanzaro},
  \bibinfo{person}{Jingdong Chen}, \bibinfo{person}{Mike Chrzanowski},
  \bibinfo{person}{Adam Coates}, \bibinfo{person}{Greg Diamos}, {and}
  \bibinfo{person}{others}.} \bibinfo{year}{2016}\natexlab{}.
\newblock \showarticletitle{Deep speech 2: End-to-end speech recognition in
  english and mandarin}. In \bibinfo{booktitle}{{\em International Conference
  on Machine Learning}}. \bibinfo{pages}{173--182}.
\newblock


\bibitem[\protect\citeauthoryear{Chand, Thakkar, and Ganatra}{Chand
  et~al\mbox{.}}{2012}]%
        {chand2012sequential}
\bibfield{author}{\bibinfo{person}{Chetna Chand}, \bibinfo{person}{Amit
  Thakkar}, {and} \bibinfo{person}{Amit Ganatra}.}
  \bibinfo{year}{2012}\natexlab{}.
\newblock \showarticletitle{Sequential pattern mining: Survey and current
  research challenges}.
\newblock \bibinfo{journal}{{\em International Journal of Soft Computing and
  Engineering\/}} \bibinfo{volume}{2}, \bibinfo{number}{1}
  (\bibinfo{year}{2012}), \bibinfo{pages}{185--193}.
\newblock


\bibitem[\protect\citeauthoryear{Chen, Moore, Turnbull, and Joachims}{Chen
  et~al\mbox{.}}{2012}]%
        {chen2012playlist}
\bibfield{author}{\bibinfo{person}{Shuo Chen}, \bibinfo{person}{Josh~L Moore},
  \bibinfo{person}{Douglas Turnbull}, {and} \bibinfo{person}{Thorsten
  Joachims}.} \bibinfo{year}{2012}\natexlab{}.
\newblock \showarticletitle{Playlist prediction via metric embedding}. In
  \bibinfo{booktitle}{{\em Proceedings of the 18th ACM SIGKDD International
  Conference on Knowledge Discovery and Data Mining}}. ACM,
  \bibinfo{pages}{714--722}.
\newblock


\bibitem[\protect\citeauthoryear{Cho, Merrienboer, Bahdanau, and Bengio}{Cho
  et~al\mbox{.}}{2014}]%
        {Cho2014On}
\bibfield{author}{\bibinfo{person}{Kyunghyun Cho}, \bibinfo{person}{Bart~Van
  Merrienboer}, \bibinfo{person}{Dzmitry Bahdanau}, {and}
  \bibinfo{person}{Yoshua Bengio}.} \bibinfo{year}{2014}\natexlab{}.
\newblock \showarticletitle{On the Properties of Neural Machine Translation:
  Encoder-Decoder Approaches}.
\newblock \bibinfo{journal}{{\em Computer Science\/}} (\bibinfo{year}{2014}).
\newblock


\bibitem[\protect\citeauthoryear{Cooper and Shallice}{Cooper and
  Shallice}{2006}]%
        {cooper2006hierarchical}
\bibfield{author}{\bibinfo{person}{R.~P. Cooper} {and} \bibinfo{person}{T
  Shallice}.} \bibinfo{year}{2006}\natexlab{}.
\newblock \showarticletitle{Hierarchical schemas and goals in the control of
  sequential behavior}.
\newblock \bibinfo{journal}{{\em Psychological Review\/}}
  \bibinfo{volume}{113}, \bibinfo{number}{4} (\bibinfo{year}{2006}),
  \bibinfo{pages}{887}.
\newblock


\bibitem[\protect\citeauthoryear{Davidson, Liebald, Liu, Nandy, Van~Vleet,
  Gargi, Gupta, He, Lambert, Livingston, et~al\mbox{.}}{Davidson
  et~al\mbox{.}}{2010}]%
        {davidson2010youtube}
\bibfield{author}{\bibinfo{person}{James Davidson}, \bibinfo{person}{Benjamin
  Liebald}, \bibinfo{person}{Junning Liu}, \bibinfo{person}{Palash Nandy},
  \bibinfo{person}{Taylor Van~Vleet}, \bibinfo{person}{Ullas Gargi},
  \bibinfo{person}{Sujoy Gupta}, \bibinfo{person}{Yu He}, \bibinfo{person}{Mike
  Lambert}, \bibinfo{person}{Blake Livingston}, {and}
  \bibinfo{person}{others}.} \bibinfo{year}{2010}\natexlab{}.
\newblock \showarticletitle{The YouTube video recommendation system}. In
  \bibinfo{booktitle}{{\em Proceedings of the 4th ACM Conference on Recommender
  Systems}}. ACM, \bibinfo{pages}{293--296}.
\newblock


\bibitem[\protect\citeauthoryear{Elkahky, Song, and He}{Elkahky
  et~al\mbox{.}}{2015}]%
        {Elkahky2015A}
\bibfield{author}{\bibinfo{person}{Ali~Mamdouh Elkahky}, \bibinfo{person}{Yang
  Song}, {and} \bibinfo{person}{Xiaodong He}.} \bibinfo{year}{2015}\natexlab{}.
\newblock \showarticletitle{A Multi-View Deep Learning Approach for Cross
  Domain User Modeling in Recommendation Systems}. In \bibinfo{booktitle}{{\em
  International Conference on World Wide Web}}. \bibinfo{pages}{278--288}.
\newblock


\bibitem[\protect\citeauthoryear{Fulkerson}{Fulkerson}{1995}]%
        {Bill1995Machine}
\bibfield{author}{\bibinfo{person}{Bill Fulkerson}.}
  \bibinfo{year}{1995}\natexlab{}.
\newblock \showarticletitle{Machine Learning, Neural and Statistical
  Classification}.
\newblock \bibinfo{journal}{{\em Technometrics\/}} \bibinfo{volume}{37},
  \bibinfo{number}{4} (\bibinfo{year}{1995}), \bibinfo{pages}{459--459}.
\newblock


\bibitem[\protect\citeauthoryear{Graves, Mohamed, and Hinton}{Graves
  et~al\mbox{.}}{2013}]%
        {graves2013speech}
\bibfield{author}{\bibinfo{person}{Alex Graves}, \bibinfo{person}{Abdel-rahman
  Mohamed}, {and} \bibinfo{person}{Geoffrey Hinton}.}
  \bibinfo{year}{2013}\natexlab{}.
\newblock \showarticletitle{Speech recognition with deep recurrent neural
  networks}. In \bibinfo{booktitle}{{\em Acoustics, Speech and Signal
  Processing (ICASSP), IEEE International Conference on}}. IEEE,
  \bibinfo{pages}{6645--6649}.
\newblock


\bibitem[\protect\citeauthoryear{He, Zhang, Ren, and Sun}{He
  et~al\mbox{.}}{2016}]%
        {he2016deep}
\bibfield{author}{\bibinfo{person}{Kaiming He}, \bibinfo{person}{Xiangyu
  Zhang}, \bibinfo{person}{Shaoqing Ren}, {and} \bibinfo{person}{Jian Sun}.}
  \bibinfo{year}{2016}\natexlab{}.
\newblock \showarticletitle{Deep residual learning for image recognition}. In
  \bibinfo{booktitle}{{\em Proceedings of the IEEE Conference on Computer
  Vision and Pattern Recognition}}. \bibinfo{pages}{770--778}.
\newblock


\bibitem[\protect\citeauthoryear{Hidasi, Karatzoglou, Baltrunas, and
  Tikk}{Hidasi et~al\mbox{.}}{2015}]%
        {hidasi2015session}
\bibfield{author}{\bibinfo{person}{Balázs Hidasi}, \bibinfo{person}{Alexandros
  Karatzoglou}, \bibinfo{person}{Linas Baltrunas}, {and}
  \bibinfo{person}{Domonkos Tikk}.} \bibinfo{year}{2015}\natexlab{}.
\newblock \showarticletitle{Session-based Recommendations with Recurrent Neural
  Networks}.
\newblock \bibinfo{journal}{{\em Computer Science\/}} (\bibinfo{year}{2015}).
\newblock


\bibitem[\protect\citeauthoryear{Hidasi and Tikk}{Hidasi and Tikk}{2012}]%
        {hidasi2012fast}
\bibfield{author}{\bibinfo{person}{Bal{\'a}zs Hidasi} {and}
  \bibinfo{person}{Domonkos Tikk}.} \bibinfo{year}{2012}\natexlab{}.
\newblock \showarticletitle{Fast ALS-based tensor factorization for
  context-aware recommendation from implicit feedback}.
\newblock \bibinfo{journal}{{\em Machine Learning and Knowledge Discovery in
  Databases\/}} (\bibinfo{year}{2012}), \bibinfo{pages}{67--82}.
\newblock


\bibitem[\protect\citeauthoryear{Hochreiter and Schmidhuber}{Hochreiter and
  Schmidhuber}{2012}]%
        {Hochreiter2012Long}
\bibfield{author}{\bibinfo{person}{Sepp Hochreiter} {and}
  \bibinfo{person}{Jürgen Schmidhuber}.} \bibinfo{year}{2012}\natexlab{}.
\newblock \showarticletitle{Long Short-Term Memory}.
\newblock \bibinfo{journal}{{\em Neural Computation\/}} \bibinfo{volume}{9},
  \bibinfo{number}{8} (\bibinfo{year}{2012}), \bibinfo{pages}{1735--1780}.
\newblock


\bibitem[\protect\citeauthoryear{Kingma and Ba}{Kingma and Ba}{2014}]%
        {kingma2014adam}
\bibfield{author}{\bibinfo{person}{Diederik Kingma} {and}
  \bibinfo{person}{Jimmy Ba}.} \bibinfo{year}{2014}\natexlab{}.
\newblock \showarticletitle{Adam: A Method for Stochastic Optimization}.
\newblock \bibinfo{journal}{{\em Computer Science\/}} (\bibinfo{year}{2014}).
\newblock


\bibitem[\protect\citeauthoryear{Koren, Bell, and Volinsky}{Koren
  et~al\mbox{.}}{2009}]%
        {koren2009matrix}
\bibfield{author}{\bibinfo{person}{Yehuda Koren}, \bibinfo{person}{Robert
  Bell}, {and} \bibinfo{person}{Chris Volinsky}.}
  \bibinfo{year}{2009}\natexlab{}.
\newblock \showarticletitle{Matrix Factorization Techniques for Recommender
  Systems}.
\newblock \bibinfo{journal}{{\em Computer\/}} \bibinfo{volume}{42},
  \bibinfo{number}{8} (\bibinfo{year}{2009}), \bibinfo{pages}{30--37}.
\newblock


\bibitem[\protect\citeauthoryear{Krizhevsky, Sutskever, and Hinton}{Krizhevsky
  et~al\mbox{.}}{2012}]%
        {krizhevsky2012imagenet}
\bibfield{author}{\bibinfo{person}{Alex Krizhevsky}, \bibinfo{person}{Ilya
  Sutskever}, {and} \bibinfo{person}{Geoffrey~E Hinton}.}
  \bibinfo{year}{2012}\natexlab{}.
\newblock \showarticletitle{Imagenet classification with deep convolutional
  neural networks}. In \bibinfo{booktitle}{{\em Advances in Neural Information
  Processing Systems}}. \bibinfo{pages}{1097--1105}.
\newblock


\bibitem[\protect\citeauthoryear{Linden, Smith, and York}{Linden
  et~al\mbox{.}}{2003}]%
        {linden2003amazon}
\bibfield{author}{\bibinfo{person}{Greg Linden}, \bibinfo{person}{Brent Smith},
  {and} \bibinfo{person}{Jeremy York}.} \bibinfo{year}{2003}\natexlab{}.
\newblock \showarticletitle{Amazon. com recommendations: Item-to-item
  collaborative filtering}.
\newblock \bibinfo{journal}{{\em IEEE Internet Computing\/}}
  \bibinfo{volume}{7}, \bibinfo{number}{1} (\bibinfo{year}{2003}),
  \bibinfo{pages}{76--80}.
\newblock


\bibitem[\protect\citeauthoryear{Liu, Chen, Cai, and Yu}{Liu
  et~al\mbox{.}}{2012}]%
        {liu2012enlister}
\bibfield{author}{\bibinfo{person}{Qiwen Liu}, \bibinfo{person}{Tianjian Chen},
  \bibinfo{person}{Jing Cai}, {and} \bibinfo{person}{Dianhai Yu}.}
  \bibinfo{year}{2012}\natexlab{}.
\newblock \showarticletitle{Enlister: baidu's recommender system for the
  biggest chinese Q\&A website}. In \bibinfo{booktitle}{{\em Proceedings of the
  6th ACM Conference on Recommender Systems}}. ACM, \bibinfo{pages}{285--288}.
\newblock


\bibitem[\protect\citeauthoryear{Mikolov, Sutskever, Chen, Corrado, and
  Dean}{Mikolov et~al\mbox{.}}{2013}]%
        {mikolov2013distributed}
\bibfield{author}{\bibinfo{person}{Tomas Mikolov}, \bibinfo{person}{Ilya
  Sutskever}, \bibinfo{person}{Kai Chen}, \bibinfo{person}{Greg Corrado}, {and}
  \bibinfo{person}{Jeffrey Dean}.} \bibinfo{year}{2013}\natexlab{}.
\newblock \showarticletitle{Distributed Representations of Words and Phrases
  and their Compositionality}.
\newblock \bibinfo{journal}{{\em Advances in Neural Information Processing
  Systems\/}}  \bibinfo{volume}{26}, \bibinfo{pages}{3111--3119}.
\newblock


\bibitem[\protect\citeauthoryear{Mild and Reutterer}{Mild and
  Reutterer}{2003}]%
        {mild2003improved}
\bibfield{author}{\bibinfo{person}{Andreas Mild} {and} \bibinfo{person}{Thomas
  Reutterer}.} \bibinfo{year}{2003}\natexlab{}.
\newblock \showarticletitle{An improved collaborative filtering approach for
  predicting cross-category purchases based on binary market basket data}.
\newblock \bibinfo{journal}{{\em Journal of Retailing and Consumer Services\/}}
  \bibinfo{volume}{10}, \bibinfo{number}{3} (\bibinfo{year}{2003}),
  \bibinfo{pages}{123--133}.
\newblock


\bibitem[\protect\citeauthoryear{Mnih and Hinton}{Mnih and Hinton}{2008}]%
        {mnih2009scalable}
\bibfield{author}{\bibinfo{person}{Andriy Mnih} {and} \bibinfo{person}{Geoffrey
  Hinton}.} \bibinfo{year}{2008}\natexlab{}.
\newblock \showarticletitle{A scalable hierarchical distributed language
  model}. In \bibinfo{booktitle}{{\em Conference on Neural Information
  Processing Systems, Vancouver, British Columbia, Canada, December}}.
  \bibinfo{pages}{1081--1088}.
\newblock


\bibitem[\protect\citeauthoryear{Mobasher, Dai, Luo, and Nakagawa}{Mobasher
  et~al\mbox{.}}{2002}]%
        {mobasher2002using}
\bibfield{author}{\bibinfo{person}{Bamshad Mobasher}, \bibinfo{person}{Honghua
  Dai}, \bibinfo{person}{Tao Luo}, {and} \bibinfo{person}{Miki Nakagawa}.}
  \bibinfo{year}{2002}\natexlab{}.
\newblock \showarticletitle{Using sequential and non-sequential patterns in
  predictive web usage mining tasks}. In \bibinfo{booktitle}{{\em IEEE
  International Conference on Data Mining, 2002. ICDM 2003. Proceedings}}.
  \bibinfo{pages}{669--672}.
\newblock


\bibitem[\protect\citeauthoryear{Rendle, Freudenthaler, Gantner, and
  Schmidt-Thieme}{Rendle et~al\mbox{.}}{2009}]%
        {rendle2009bpr}
\bibfield{author}{\bibinfo{person}{Steffen Rendle}, \bibinfo{person}{Christoph
  Freudenthaler}, \bibinfo{person}{Zeno Gantner}, {and} \bibinfo{person}{Lars
  Schmidt-Thieme}.} \bibinfo{year}{2009}\natexlab{}.
\newblock \showarticletitle{BPR: Bayesian personalized ranking from implicit
  feedback}. In \bibinfo{booktitle}{{\em Proceedings of the 25th Conference on
  Uncertainty in Artificial Intelligence}}. AUAI Press,
  \bibinfo{pages}{452--461}.
\newblock


\bibitem[\protect\citeauthoryear{Rendle, Freudenthaler, and
  Schmidt-Thieme}{Rendle et~al\mbox{.}}{2010}]%
        {rendle2010factorizing}
\bibfield{author}{\bibinfo{person}{Steffen Rendle}, \bibinfo{person}{Christoph
  Freudenthaler}, {and} \bibinfo{person}{Lars Schmidt-Thieme}.}
  \bibinfo{year}{2010}\natexlab{}.
\newblock \showarticletitle{Factorizing personalized markov chains for
  next-basket recommendation}. In \bibinfo{booktitle}{{\em Proceedings of the
  19th International Conference on World Wide Web}}. ACM,
  \bibinfo{pages}{811--820}.
\newblock


\bibitem[\protect\citeauthoryear{Resnick and Varian}{Resnick and
  Varian}{1997}]%
        {resnick1997recommender}
\bibfield{author}{\bibinfo{person}{Paul Resnick} {and} \bibinfo{person}{Hal~R
  Varian}.} \bibinfo{year}{1997}\natexlab{}.
\newblock \showarticletitle{Recommender systems}.
\newblock \bibinfo{journal}{{\it Commun. ACM}} \bibinfo{volume}{40},
  \bibinfo{number}{3} (\bibinfo{year}{1997}), \bibinfo{pages}{56--58}.
\newblock


\bibitem[\protect\citeauthoryear{Salakhutdinov, Mnih, and Hinton}{Salakhutdinov
  et~al\mbox{.}}{2007}]%
        {Salakhutdinov2007Restricted}
\bibfield{author}{\bibinfo{person}{Ruslan Salakhutdinov},
  \bibinfo{person}{Andriy Mnih}, {and} \bibinfo{person}{Geoffrey Hinton}.}
  \bibinfo{year}{2007}\natexlab{}.
\newblock \showarticletitle{Restricted Boltzmann machines for collaborative
  filtering}. In \bibinfo{booktitle}{{\em International Conference on Machine
  Learning}}. \bibinfo{pages}{791--798}.
\newblock


\bibitem[\protect\citeauthoryear{Sarwar, Karypis, Konstan, and Riedl}{Sarwar
  et~al\mbox{.}}{2001}]%
        {sarwar2001item}
\bibfield{author}{\bibinfo{person}{Badrul Sarwar}, \bibinfo{person}{George
  Karypis}, \bibinfo{person}{Joseph Konstan}, {and} \bibinfo{person}{John
  Riedl}.} \bibinfo{year}{2001}\natexlab{}.
\newblock \showarticletitle{Item-based collaborative filtering recommendation
  algorithms}. In \bibinfo{booktitle}{{\em Proceedings of the 10th
  International Conference on World Wide Web}}. ACM, \bibinfo{pages}{285--295}.
\newblock


\bibitem[\protect\citeauthoryear{Schafer, Konstan, and Riedl}{Schafer
  et~al\mbox{.}}{1999}]%
        {schafer1999recommender}
\bibfield{author}{\bibinfo{person}{J~Ben Schafer}, \bibinfo{person}{Joseph
  Konstan}, {and} \bibinfo{person}{John Riedl}.}
  \bibinfo{year}{1999}\natexlab{}.
\newblock \showarticletitle{Recommender systems in e-commerce}. In
  \bibinfo{booktitle}{{\em Proceedings of the 1st ACM Conference on Electronic
  Commerce}}. ACM, \bibinfo{pages}{158--166}.
\newblock


\bibitem[\protect\citeauthoryear{Sedhain, Menon, Sanner, and Xie}{Sedhain
  et~al\mbox{.}}{2015}]%
        {sedhain2015autorec}
\bibfield{author}{\bibinfo{person}{Suvash Sedhain},
  \bibinfo{person}{Aditya~Krishna Menon}, \bibinfo{person}{Scott Sanner}, {and}
  \bibinfo{person}{Lexing Xie}.} \bibinfo{year}{2015}\natexlab{}.
\newblock \showarticletitle{Autorec: Autoencoders meet collaborative
  filtering}. In \bibinfo{booktitle}{{\em Proceedings of the 24th International
  Conference on World Wide Web}}. ACM, \bibinfo{pages}{111--112}.
\newblock


\bibitem[\protect\citeauthoryear{Shang, Lu, and Li}{Shang
  et~al\mbox{.}}{2015}]%
        {Shang2015Neural}
\bibfield{author}{\bibinfo{person}{Lifeng Shang}, \bibinfo{person}{Zhengdong
  Lu}, {and} \bibinfo{person}{Hang Li}.} \bibinfo{year}{2015}\natexlab{}.
\newblock \showarticletitle{Neural Responding Machine for Short-Text
  Conversation}.
\newblock \bibinfo{journal}{{\em Computer Science\/}} (\bibinfo{year}{2015}).
\newblock


\bibitem[\protect\citeauthoryear{Shani, Brafman, and Heckerman}{Shani
  et~al\mbox{.}}{2002}]%
        {shani2005mdp}
\bibfield{author}{\bibinfo{person}{Guy Shani}, \bibinfo{person}{Ronen~I
  Brafman}, {and} \bibinfo{person}{David Heckerman}.}
  \bibinfo{year}{2002}\natexlab{}.
\newblock \showarticletitle{An MDP-based recommender system}.
\newblock  (\bibinfo{year}{2002}), \bibinfo{pages}{453--460}.
\newblock


\bibitem[\protect\citeauthoryear{Socher, Lin, Ng, and Manning}{Socher
  et~al\mbox{.}}{2011}]%
        {socher2011parsing}
\bibfield{author}{\bibinfo{person}{Richard Socher}, \bibinfo{person}{Chiung~Yu
  Lin}, \bibinfo{person}{Andrew~Y. Ng}, {and} \bibinfo{person}{Christopher~D.
  Manning}.} \bibinfo{year}{2011}\natexlab{}.
\newblock \showarticletitle{Parsing Natural Scenes and Natural Language with
  Recursive Neural Networks}. In \bibinfo{booktitle}{{\em International
  Conference on Machine Learning (ICML)}}. \bibinfo{pages}{129--136}.
\newblock


\bibitem[\protect\citeauthoryear{Su and Khoshgoftaar}{Su and
  Khoshgoftaar}{2009}]%
        {su2009survey}
\bibfield{author}{\bibinfo{person}{Xiaoyuan Su} {and} \bibinfo{person}{Taghi~M
  Khoshgoftaar}.} \bibinfo{year}{2009}\natexlab{}.
\newblock \showarticletitle{A survey of collaborative filtering techniques}.
\newblock \bibinfo{journal}{{\em Advances in Artificial Intelligence\/}}
  (\bibinfo{year}{2009}), \bibinfo{pages}{4}.
\newblock


\bibitem[\protect\citeauthoryear{Tan, Xu, and Liu}{Tan et~al\mbox{.}}{2016}]%
        {tan2016improved}
\bibfield{author}{\bibinfo{person}{Yong~Kiam Tan}, \bibinfo{person}{Xinxing
  Xu}, {and} \bibinfo{person}{Yong Liu}.} \bibinfo{year}{2016}\natexlab{}.
\newblock \showarticletitle{Improved recurrent neural networks for
  session-based recommendations}. In \bibinfo{booktitle}{{\em Proceedings of
  the 1st Workshop on Deep Learning for Recommender Systems}}. ACM,
  \bibinfo{pages}{17--22}.
\newblock


\bibitem[\protect\citeauthoryear{Wang, Guo, Lan, Xu, Wan, and Cheng}{Wang
  et~al\mbox{.}}{2015}]%
        {Wang2015Learning}
\bibfield{author}{\bibinfo{person}{Pengfei Wang}, \bibinfo{person}{Jiafeng
  Guo}, \bibinfo{person}{Yanyan Lan}, \bibinfo{person}{Jun Xu},
  \bibinfo{person}{Shengxian Wan}, {and} \bibinfo{person}{Xueqi Cheng}.}
  \bibinfo{year}{2015}\natexlab{}.
\newblock \showarticletitle{Learning Hierarchical Representation Model for
  NextBasket Recommendation}. In \bibinfo{booktitle}{{\em SIGIR}}.
  \bibinfo{pages}{403--412}.
\newblock


\bibitem[\protect\citeauthoryear{Wu, Dubois, Zheng, and Ester}{Wu
  et~al\mbox{.}}{2016}]%
        {Wu2016Collaborative}
\bibfield{author}{\bibinfo{person}{Yao Wu}, \bibinfo{person}{Christopher
  Dubois}, \bibinfo{person}{Alice~X Zheng}, {and} \bibinfo{person}{Martin
  Ester}.} \bibinfo{year}{2016}\natexlab{}.
\newblock \showarticletitle{Collaborative Denoising Auto-Encoders for Top-N
  Recommender Systems}. In \bibinfo{booktitle}{{\em ACM International
  Conference on Web Search and Data Mining}}. \bibinfo{pages}{153--162}.
\newblock


\bibitem[\protect\citeauthoryear{Yap, Li, and Philip}{Yap
  et~al\mbox{.}}{2012}]%
        {yap2012effective}
\bibfield{author}{\bibinfo{person}{Ghim-Eng Yap}, \bibinfo{person}{Xiao-Li Li},
  {and} \bibinfo{person}{S~Yu Philip}.} \bibinfo{year}{2012}\natexlab{}.
\newblock \showarticletitle{Effective next-items recommendation via
  personalized sequential pattern mining}. In \bibinfo{booktitle}{{\em
  International Conference on Database Systems for Advanced Applications}}.
  Springer, \bibinfo{pages}{48--64}.
\newblock


\bibitem[\protect\citeauthoryear{Zhang, Dai, Xu, Feng, Wang, Bian, Wang, and
  Liu}{Zhang et~al\mbox{.}}{2014}]%
        {Zhang2014Sequential}
\bibfield{author}{\bibinfo{person}{Yuyu Zhang}, \bibinfo{person}{Hanjun Dai},
  \bibinfo{person}{Chang Xu}, \bibinfo{person}{Jun Feng},
  \bibinfo{person}{Taifeng Wang}, \bibinfo{person}{Jiang Bian},
  \bibinfo{person}{Bin Wang}, {and} \bibinfo{person}{Tie~Yan Liu}.}
  \bibinfo{year}{2014}\natexlab{}.
\newblock \showarticletitle{Sequential click prediction for sponsored search
  with recurrent neural networks}.
\newblock  (\bibinfo{year}{2014}), \bibinfo{pages}{1369--1375}.
\newblock


\bibitem[\protect\citeauthoryear{Zimdars, Chickering, and Meek}{Zimdars
  et~al\mbox{.}}{2001}]%
        {zimdars2001using}
\bibfield{author}{\bibinfo{person}{Andrew Zimdars},
  \bibinfo{person}{David~Maxwell Chickering}, {and}
  \bibinfo{person}{Christopher Meek}.} \bibinfo{year}{2001}\natexlab{}.
\newblock \showarticletitle{Using temporal data for making recommendations}. In
  \bibinfo{booktitle}{{\em Proceedings of the 7th conference on Uncertainty in
  Artificial Intelligence}}. Morgan Kaufmann Publishers Inc.,
  \bibinfo{pages}{580--588}.
\newblock


\end{thebibliography}

\end{document}